\pdfoutput=1
 \newif\iffigs
  \figsfalse
\documentclass[12pt]{article}
\usepackage{colordvi}
\usepackage{latexsym,amssymb}
\iffigs
    \usepackage{graphicx}
    \input{epsf}
\else
\fi
\newcommand{\ft}[2]{{\textstyle\frac{#1}{#2}}}
\def\tilde{\widetilde}

\newsavebox{\uuunit}
\sbox{\uuunit}
                             {\setlength{\unitlength}{0.825em}
                                      \begin{picture}(0.6,0.7)
                                                                  \thinlines
                                                                  \put(0,0){\line(1,0){0.5}}
                                                                  \put(0.15,0){\line(0,1){0.7}}
                                                                  \put(0.35,0){\line(0,1){0.8}}
                                                                 \multiput(0.3,0.8)(-0.04,-0.02){10}{\rule{0.5pt}{0.5pt}}
                                      \end {picture}}
\newcommand {\unity}{\mathord{\!\usebox{\uuunit}}}
\makeatletter \@addtoreset{equation}{section} \makeatother

\newcommand{\OSp}{\mathop{\rm {}OSp}}

\def\bfone{\relax{\rm 1\kern-.35em 1}}

\def\tilde{\widetilde}

\def\tn010{\tilde{N}^{0,1,0}}

\newcommand{\so}{\mathfrak{so}}
\newcommand{\su}{\mathfrak{su}}

\newcommand{\sym}{\mathfrak{sp}}
\newcommand{\uu}{\mathfrak{u}}
\newcommand{\osp}{\mathfrak{osp}}
\usepackage{ifpdf}
\ifpdf
   \pdfoutput=1
   \pdftrue
  \usepackage[pdftex]{hyperref}
 \else
   \usepackage{cite}
\fi
\begin{document}
\begin{titlepage}
\begin{flushright}
DFTT/xxx\\
DISTA-2007 \\
hep-th/yymmnnn
\end{flushright}
\vskip 1.5cm
\begin{center}
{\LARGE \bf
 Constrained Supermanifolds for\\
 \vskip 0.2cm
  $AdS$ M-Theory
  Backgrounds$^\dagger$
} \\
 \vfill
{\large Pietro Fr{\'e}$^1$ and Pietro Antonio Grassi$^2$,} \\
\vfill {
$^1$ Dipartimento di Fisica Teorica, Universit{\`a} di Torino, \\
$\&$ INFN -
Sezione di Torino\\
via P. Giuria 1, I-10125 Torino, Italy\\
\vskip .3cm
$^2$ { Centro Studi e Ricerche E. Fermi,
Compendio Viminale, I-00184, Roma, Italy,}\\
{ DISTA, Universit\`a del Piemonte Orientale, }\\
{ Via Bellini 25/G,  Alessandria, 15100, Italy
$\&$ INFN - Sezione di
Torino}
}
\end{center}
\vfill
\begin{abstract}
A long standing problem is the supergauge completion of
$\mathrm{AdS}_4 \, \times \, \left( \mathcal{G/H}\right) _7$ or
$\mathrm{AdS}_5 \, \times \, \left( \mathcal{G/H}\right) _5$ backgrounds
which preserve less then maximal supersymmetry.
In parallel with the supersolvable realization of the $\mathrm{AdS_4} \times \mathbb{S}^7$ background based on
$\kappa$-symmetry, we develop a technique which amounts to solving the above-mentioned problem in a way useful for pure spinor
quantization for supermembranes and superstrings. Instead of gauge fixing some
of the superspace coordinates to zero, we impose an additional constraint on
them reproducing the simplifications of the supersolvable representations.
The constraints are quadratic, homogeneous, $\mathrm{Sp(4,\mathbb{R})}$-covariant, and consistent from
the quantum point of view in the pure spinor approach. Here we provide the
geometrical solution which, in a subsequent work, will be applied
to the membrane and the superstring sigma models.
\end{abstract}
\vfill
\vspace{1.5cm}
\vspace{2mm} \vfill \hrule width 3.cm {\footnotesize $^ \dagger $
This work is supported in part by the European Union RTN contract
MRTN-CT-2004-005104 and by the Italian Ministry of University (MIUR) under
contracts PRIN 2005-024045 and PRIN 2005-023102}
\end{titlepage}
\tableofcontents
\section{Introduction}
\label{introibo}
One of the most interesting progress in the theory of supermembranes
is the quantization by using the pure spinor technique \cite{berko_mem}.
It provides a quantum model (interacting) where the kappa symmetry is
gauge fixed and a BRST is provided. Using the BRST operators one can compute the cohomology and the spectrum. Unfortunately, the interacting
worldvolume action does not allow a simple analysis of the complete spectrum
and only the massless sector can be studied by using the target space symmetries.
Nevertheless, the main advantage is a complete superspace description
of the theory in terms of vielbeins, gravitinos and the superfield generalization
of the $3$-form of 11-dimensional supergravity \cite{cremmerjulia,fredauria11}. 
Recently in \cite{antonpietro},
we have shown that there is a deep relation between the pure spinor BRST symmetry 
\cite{Berkovits:2000fe,Berkovits:2001ue} and the
Free Differential Algebra of 11 supergravity and we have used these facts to
obtain a complete algebraic derivation of the BRST symmetry and
of the symmetries of the model. The resulting action has a manifest
supersymmetry and it depends on the supergravity background superfields.
Those superfields are obtained from the FDA by gauge completing the superfield starting from a given bosonic background which satisfies the equations of motion.
(We have to remind the reader that the FDA's for
11-dimensional supergravity discussed in sec. \ref{superFDA} imply
the equations of motion).
\par
However, to solve the FDA for a given
background is not a trivial task and the complete superfield is needed in order
to compute amplitudes in presence of a given background. In practice one
needs a superfield only up to a certain power in the fermionic coordinates.
The reason is that the coefficients of higher powers are simply ordinary derivatives of the  lowest components and they do no carry new information.
Nevertheless, those coefficients enter the computation of amplitudes and
we need a  method to reconstruct a complete superfield in terms
of the bosonic solution. There are on the market several techniques, see
for example \cite{ooguri, mio-tamassia, tsimpis, vanhove} just to quote
some of them adapted to our problem. These techniques start from a
very general setting and they provide an iterative reconstruction method,
which unfortunately hides completely the geometry behind the solution.
We take a different perspective: we start from a  solution with some supersymmetries
(in our case, from the 4-dimensional point of view we take the
supersymmetric models with ${\mathcal{N}} =8,3,1$) and some relevant isometries and we try to build a
complete superfield solution respecting these symmetries. The rheonomic parametrizations of FDA.s are
integrable by construction and
the consistency conditions are just the equations of motion \cite{fredauria11}.
Therefore we need to start from an on-shell background solution and we are guaranteed that the solution exists.
The best way to find complete solutions of the FDA is terms
of a super-Lie algebra and of its Maurer-Cartan forms. As will be discussed in
next sections, one starts from the Killing spinor of the bosonic solution and
he reconstructs the gravitino fields by ``pairing'' the Killing
spinors of the bosonic submanifold
with fermionic Maurer-Cartan forms of the underlying algebra. Then,
by inserting the gravitino field in the FDA and using the relations between
the Maurer-Cartan forms dictated by the Lie superalgebra, one finds that
the gravitinos satisfy their own equations. In the same way
one can modfify the bosonic supervielbein by adding
the bosonic Maurer-Cartan forms and, by inserting it into the FDA equations,
one finds all correct pieces. This technique permits a direct complete
solution of the gauge completion only for supergroups or supercosets. It does not
work that simply in the case of less supersymmetry of the background, and
some modifications are needed.
\par
First, one needs to study the obstruction that prevents one from  getting
a complete solution as a supergroup or a supercoset. This is
parameterized by the Weyl tensor which is obtained by commuting
two covariant derivatives. Second, one finds that some of the structures
of the supercoset technique can still be used. For example, one can
organize the fermionic coordinates in two sectors: 1) those related to the
linear realization of supersymmetry (the unbroken supersymmetries) and
2) the remaining set related to the broken supersymmetries, and the most convenient method seems to follow very closely the supercoset solution. We
assume that the fermionic coordinates are organized according to
a pure fermionic supercoset and we construct the gravitinos by pairing the
Killing spinors and some other spinor (needed to span a complete
basis of sections of the spinor bundle over the bosonic submanifold) with the Maurer-Cartan forms. The violation of the FDA can be
compensated by adding to the gravitions and to other superfield additional
pieces. These pieces can be taken automatically into account, by
promoting the Maurer-Cartan forms to \textit{gauged Maurer-Cartan forms}.
This yields an additional term in the vielbein equation which can be reabsorbed
into a redefintion of the spin connection. In this way the procedure can be iterated (even if it will not be pursued here further)
and one lands with a complete superfield construction.
\par
Fortunately, there is an interesting alternative to the iterative solution.
This procedure has been developed in \cite{Dall'Agata:1998wz,Pesando:1998fv} and used in several applications (see for example \cite{Kallosh:1998ji})
and it is based on the supersolvable realization of the supercoset $\mathrm{Osp(8|4)}/\mathrm{SO(1,3)} \times \mathrm{Sp(4,\mathbb{R})}$ in the case of M-theory
and of $\mathrm{SU(2,2|4)}/\mathrm{SO(1,4)} \times \mathrm{SO(5)}$ for the superstring. Using the
$\kappa$-symmetry one can gauge some coordinates of the superspace
to zero and write the Maurer-Cartan equations only in terms of the reduced superspace. This has the advantage to fix the gauge symmetry and to
simplify the Maurer-Cartan forms drastically. Specifically it turns out that after this gauge fixing,  they are just quadratic
in the $\theta$-coordinates. In this way, the problem of resumming the
complete dependence of the fermionic coordinates is avoided and
the gauged Maurer-Cartan equations already suffice to solve the problem of the
gauge completion. Indeed, only a remaining additional piece of contorsion
must be added in order to compensate the non-vanishing of the Weyl tensor.
\par
This for what concerns the models with $\kappa$-symmetry where the
gauge completion can be provided. However, we notice that the same
simplification can be achieved by imposing a constraint on the
fermionic coordinates. In the case of $\mathrm{Osp(8|4)}/\mathrm{SO(1,3)} \times \mathrm{Sp(4,\mathbb{R})}$
is
\begin{equation}
\Theta^{x}_{A} \epsilon_{xy} \Theta^{y}_{B} =0\,.
\end{equation}
Here the indices $A,B$ run over $1, \dots,8$ and the indices $x,y$ over
$1,\dots,4$. The equation is symmetric in the $\mathrm{SO(8)}$ indices, it is
homogeneous of degree two in the scaling of $\Theta$'s, is quadratic and
it is $\mathrm{Sp(4,\mathbb{R})}$ covariant which means that it does not spoil the isometries
of the $\mathrm{AdS_{4}}$ manifold. It will be shown in the text that these constraints yield the same
simplification of the supersolvable realization of the supercoset, and
in particular the $\kappa$-symmetry gauge adopted in
\cite{Dall'Agata:1998wz,Pesando:1998fv} is a solution of these
new constraints. However, in the case of Green-Schwarz type of models these constraints are
not consistent with the canonical quantization of the model. This
is due to fact that in the canonical quantization the $\Theta$'s satisfy a Clifford
algebra and the above constraints are not consistent. On the other side,
using the pure spinor formalism the commutation relations
among $\Theta$'s vanish (they have a non-vanishing commutation relations
with the conjugate momenta, see for example \cite{antonpietro}) and the constraints are consistent. In addition, they have the same dignity of the
pure spinor constraints and they can be treated on the same footing. (We
also mention that quadratic constraints for the supercoordinates appeared also
in \cite{berkovits-maldacena,Berkovits:2002rd,Berkovits:2005bt} and in \cite{witten-twistors}. In \cite{Grassi:2004cz}, which is based on pure spinor 
formulation of BRST symmetry \cite{Grassi:2001ug,Grassi:2002tz}, quadratic 
constraints for anticommuting ghosts have been discussed.)
\par
In this way, we can use the advantages of the supersolvable
description of the background in order to derive pure spinor
sigma models for supermembrane and superstrings. This can be useful
for maximal supersymmetric background and for less than maximal supersymmetric spaces.
\par
The paper is organized as follows. In sec. 2 and sec. 3, we give some details about
compactifications of the bosonic background of the form $\mathrm{AdS_{4}} \times {\mathcal{G}/\mathcal{H}}$, free
differential algebras and some notations.
In sec. 4, we recall the geometry of the spinor bundle and the holonomy tensor.
In sec. 5 we discuss some property of the supergroup $\mathrm{Osp(\mathcal{N}|4)}$ and its
Maurer-Cartan forms. Finally, we discuss the gauging and we discuss the solution to the first order. Then, we consider two examples in sec. 9.
Some additional material is contained in the Appendices.
\section{The super FDA of M theory}
\label{superFDA}
Let us begin by  writing the complete set of curvatures defining  the complete FDA of $D=11$
M-theory. As usual this FDA is the semidirect sum of a minimal
algebra with a contractible algebra:
\begin{equation}
  \mathbb{A}=\mathbb{M}\biguplus\mathbb{C}
\label{completealg}
\end{equation}
the curvatures being the contractible generators $\mathbb{C}$. By setting them to zero we  retrieve,
according to Sullivan's first theorem, the minimal
algebra $\mathbb{M}$. This latter, according to Sullivan's second
theorem, is explained in terms of cohomology of the
 super Lie subalgebra $\mathbb{G} \subset \mathbb{M}$, spanned by the
 $1$--forms. In this case  $\mathbb{G}$ is just
  the $D=11$ superPoincar\'e algebra spanned by the following $1$--forms:
\begin{enumerate}
  \item the vielbein $V^{\underline{a}}$
  \item the spin connection $\omega^{\underline{ab}}$
  \item the gravitino $\Psi$
\end{enumerate}
where the underlined indices $\underline{a},\underline{b},\dots$ run
on eleven values and are vector indices of $\mathrm{SO(1,10)}$. The
gravitino $\Psi$ is a fermionic one-form (hence commuting) assigned
to the $32$-component Majorana spinor representation of $\mathrm{SO(1,10)}$:
\begin{equation}
  C \overline{\Psi}^T = \Psi \quad ; \quad \overline{\Psi} \equiv
  \Psi^\dagger \, \Gamma_0
\label{Majocondo}
\end{equation}
\par
The higher degree generators of the minimal FDA $\mathbb{M}$ are:
\begin{enumerate}
  \item the bosonic $3$--form $\mathbf{A^{[3]}}$
  \item the bosonic $6$-form $\mathbf{A^{[6]}}$.
\end{enumerate}
The complete set of curvatures is given below (\cite{fredauria,comments}):
\begin{eqnarray}
T^{\underline{a}} & = & \mathcal{D}V^{\underline{a}}
- {\rm i} \ft 12 \, \overline{\Psi} \, \wedge \, \Gamma^{\underline{a}} \, \Psi \nonumber\\
R^{\underline{ab}} & = & d\omega^{\underline{ab}} - \omega^{\underline{ac}} \, \wedge \, \omega^{\underline{cb}}
\nonumber\\
\rho & = & \mathcal{D}\Psi \equiv d \Psi - \ft 14 \, \omega^{\underline{ab}} \, \wedge \, \Gamma_{\underline{ab}} \, \Psi\nonumber\\
\mathbf{F^{[4]}} & = & d\mathbf{A^{[3]}} - \ft 12\, \overline{\Psi} \, \wedge \, \Gamma_{\underline{ab}} \, \Psi \,
\wedge \, V^{\underline{a}} \wedge V^{\underline{b}} \nonumber\\
\mathbf{F^{[7]}} & = & d\mathbf{A^{[6]}} -15 \, \mathbf{F^{[4]}} \, \wedge \,  \mathbf{A^{[3]}} - \ft {15}{2} \,
\, V^{\underline{a}}\wedge V^{\underline{b}} \, \wedge \, {\bar \Psi}
\wedge \, \Gamma_{\underline{ab}} \, \Psi
\, \wedge \, \mathbf{A^{[3]}} \nonumber\\
\null & \null & - {\rm i}\, \ft {1}{2} \, \overline{\Psi} \, \wedge \, \Gamma_{\underline{a_1 \dots a_5}} \, \Psi \,
\wedge \, V^{\underline{a_1}} \wedge \dots \wedge V^{\underline{a_5}}
\label{FDAcompleta}
\end{eqnarray}
From their very definition, by taking a further exterior derivative
one obtains the Bianchi identities:
\begin{eqnarray}
&&\mathcal{D} R^{\underline{ab}}=0 \nonumber\\
&&\mathcal{D} T^{\underline{a}}\,
+ \,R^{\underline{a}}{}_{\underline{b}}\wedge V^{\underline{b}}+\bar{\Psi}\wedge\Gamma^{\underline{a}}\rho =
0 \nonumber\\
&&\mathcal{D}\rho + \frac{1}{4}R^{\underline{ab}}\wedge\Gamma_{\underline{ab}}\Psi=0\,,\nonumber\\
&&d\mathbf{F^{[4]}} \, - \, \bar{\Psi}\Gamma_{\underline{ab}}\, \wedge \, \rho \, \wedge V^{\underline{a}}
\, \wedge \, V^{\underline{b}}\, - \,
   \bar \Psi \, \wedge \, \Gamma_{\underline{ab}}  \Psi \, \wedge \, V^{\underline{a}}\, \wedge \,
   T^{\underline{b}}=0
\label{Bianchis}
\end{eqnarray}
The dynamical theory is
defined, according to the general constructive scheme of supersymmetric theories, by the principle
of rheonomy (see \cite{castdauriafre} ) implemented into Bianchi identities.
Indeed there is a unique rheonomic parametrization of the curvatures
which solves the
Bianchi identities and it  is the following one:
\begin{eqnarray}
T^{\underline{a}} & = & 0 \label{torsequa}\\
\mathbf{F^{[4]}} & = & F_{\underline{a_1\dots a_4}} \, V^{\underline{a_1}} \, \wedge \dots
\wedge \, V^{\underline{a_4}} \label{f4equa}\\
\mathbf{F^{[7]}} & = & \ft {1}{84} F^{\underline{a_1\dots a_4}} \, V^{\underline{b_1}} \, \wedge \dots \wedge \,
V^{\underline{b_7}} \, \epsilon_{\underline{a_1 \dots a_4 b_1 \dots b_7}} \label{f7equa}\\
\rho & = & \rho_{\underline{a_1a_2}} \,V^{\underline{a_1}} \, \wedge \, V^{\underline{a_2}} + {\rm i} \ft 13 \,
\left(\Gamma^{\underline{a_1a_2 a_3}} \Psi \, \wedge \, V^{\underline{a_4}} - \ft 1 8
\Gamma^{\underline{a_1\dots a_4 m}}\, \Psi \, \wedge \, V^{\underline{m}}
\right) \, F^{\underline{a_1 \dots a_4}} \label{rhoequa}\\
R^{\underline{ab}} & = & R^{\underline{ab}}_{\phantom{ab}\underline{cd}} \, V^{\underline{c}} \, \wedge \,
V^{\underline{d}}
+ {\rm i} \, \overline{\rho}_{\underline{mn}} \, \left( \ft 12 \Gamma^{\underline{abmnc}} - \ft 2 9
\Gamma^{\underline{mn}[\underline{a}}\, \delta^{\underline{b}]\underline{c}} + 2 \,
\Gamma^{\underline{ab}[\underline{m}} \, \delta^{\underline{n}]\underline{c}}\right) \,
\Psi \wedge V^{\underline{c}}\nonumber\\
 & &+\overline{\Psi} \wedge \, \Gamma^{\underline{mn}} \, \Psi \, F^{\underline{mnab}} +
 \ft 1{24} \overline{\Psi} \wedge \,
 \Gamma^{\underline{abc_1 \dots c_4}} \, \Psi \, F^{\underline{c_1 \dots c_4}}
\label{rheoFDA}
\end{eqnarray}
The expressions (\ref{torsequa}-\ref{rheoFDA}) satisfy the Bianchi.s provided the space--time components
of the curvatures satisfy the following constraints
\begin{eqnarray}
0 & = & \mathcal{D}_{\underline{m}} F^{\underline{mc_1 c_2 c_3}} \, +
\, \ft 1{96} \, \epsilon^{\underline{c_1c_2c_3 a_1 a_8}} \, F_{\underline{a_1 \dots a_4}}
\, F_{\underline{a_5 \dots a_8}}  \nonumber\\
0 & = & \Gamma^{\underline{abc}} \, \rho_{\underline{bc}} \nonumber\\
R^{\underline{am}}_{\phantom{\underline{bm}}\underline{cm}} & = & 6
\, F^{\underline{ac_1c_2c_3}} \,F^{\underline{bc_1c_2c_3}} -
\, \ft 12 \, \delta^{\underline{a}}_{\underline{b}} \, F^{\underline{c_1 \dots c_4}} \,F^{\underline{c_1 \dots c_4}}
\label{fieldeque}
\end{eqnarray}
which are the  space--time field equations.
\subsection{Other relevant implications of the Bianchi identities}
\label{brokensusy}
For later use it is convenient to rewrite  eq.s (\ref{rheoFDA}) in a slightly more compact form,
namely:
\begin{eqnarray}
T^{\underline{a}} &\equiv & 0\,,\nonumber\\
R^{\underline{ab}} &\equiv & R^{\underline{ab}}{}_{\underline{mn}}\, V^{\underline{m}}\, \wedge
\, V^{\underline{n}}\, +\,
\bar{\Theta}^{\underline{c}\, |\,\underline{ab}} \,
\Psi\, \wedge \, V_{\underline{c}}\, + \, \overline{\Psi}\, \wedge \, S^{\underline{ab}}\, \Psi\,,\nonumber \\
\rho &\equiv & \rho_{\underline{ab}} \, V^{\underline{a}}\, \wedge \,
V^{\underline{b}} \, + \, F_{\underline{a}}\, \Psi\, \wedge \, V^{\underline{a}}\,,\nonumber\\
\mathbf{F^{[4]}} &\equiv & F_{\underline{b}_1\dots \underline{b}_4}V^{\underline{b}_1}\wedge \dots \wedge
V^{\underline{b}_4}\,.
\label{rheoConstr}
\end{eqnarray}
where we have defined the following spinor and the following matrices:
\begin{eqnarray}
\bar{\Theta}^{\underline{c}\, |\,\underline{ab}} &=& {\rm i} \, \overline{\rho}_{\underline{mn}} \,
\left( \ft 12 \Gamma^{\underline{abmnc}} - \ft 2 9
\Gamma^{\underline{mn}[\underline{a}}\, \delta^{\underline{b}]\underline{c}} + 2 \,
\Gamma^{\underline{ab}[\underline{m}} \, \delta^{\underline{n}]\underline{c}}\right)
\,\,\nonumber\\
& = & - {\rm i} \, \overline{\rho}_{\underline{ab}} \,
\Gamma_{\underline{c}}\, + \, 2 \,{\rm i} \, \overline{\rho}_{\underline{c[a}} \,
\Gamma_{\underline{b]}}\,\label{defbartheta}\\
F_{\underline{a }} & = & T_{\underline{a}}{}^{\underline{b_1b_2b_3b_4}}
F_{\underline{b_1b_2b_3b_4}}\,,
\label{defHCF}\\
S^{\underline{ab}} & = & F^{\underline{abcd}}\Gamma _{\underline{cd}}\, + \, \ft1{24}
F_{\underline{c_1\ldots c_4}}\Gamma^{\underline{abc_1\ldots c_4}}\,,
\label{defSCF}
\end{eqnarray}
and where where we have used the following abbreviation as in \cite{CFD11}:
\begin{eqnarray}
T_{\underline{a}}{}^{\underline{b_1b_2b_3b_4}}&=& - \ft{\rm i}{24}
\left(\Gamma^{\underline{b_1b_2b_3b_4}}{}_{\underline{a}} \, + \, 8
\,
\delta_{\underline{a}}{}^{[\underline{b_1}} \Gamma^{\underline{b_2b_3b_4}]}\right)
\label{defT}\,.
\end{eqnarray}
In eq.(\ref{defbartheta}) the equality of the first with the second
line follows from the gravitino field equation, namely the second of eq.s
(\ref{fieldeque}). This latter implies that the spinor tensor
$\rho_{\underline{ab}}$ is an irreducible representation $\left(\ft 32 \, ,
\, \ft 32 \, ,  \,\ft 12 \, , \, \ft 12 \, , \,\ft 12 \,\right)$ of
$\mathrm{SO(1,10)}$, \textit{i.e}:
\begin{equation}
 \Gamma^{\underline{m}} \,
\rho_{\underline{am}}=0 \label{irreducrho}
\end{equation}
As we demonstrate later on the most important relations to be
extracted from Bianchi identities, besides the rheonomic
parametrization, concerns the spinor derivatives of the curvature
superfield. This latter is determined from the expansion of the inner
components of the 4--form field strength
$F_{\underline{a_1\dots a_4}}$. From the last of eq.s
(\ref{Bianchis}) we obtain:
\begin{eqnarray}
\mathcal{D}_{\underline{\alpha}} F_{\underline{abcd}} &=& (\Gamma_{[\underline{ab}}
\rho_{\underline{cd}]})_{\underline{\alpha}}\,,
\label{onespinorDF}
\end{eqnarray}
where the spinor derivative is normalized according to the
definition:
\begin{equation}
  \mathcal{D}\, F_{\underline{abcd}} \,  \equiv \,  \overline{\Psi}^{\underline{\alpha}} \,
\mathcal{D}_{\underline{\alpha}} F_{\underline{abcd}} \, + \,
V^{\underline{m}} \, \mathcal{D}_{\underline{m}} \,  F_{\underline{abcd}}
\label{normaspin}
\end{equation}
This shows that the gravitino field strength appears at first order
in the $\theta$-expansion of the curvature superfield. Next we
consider the spinor derivative of the gravitino field strength
itself. Using the normalization which streams from the following
definition:
\begin{equation}
  \mathcal{D}\, \rho_{\underline{ab}} =\mathcal{D}_{\underline{c}}\,\rho_{\underline{ab}}
  \, V^{\underline{c}} \, + \, K_{\underline{ab}} \, \Psi
\label{festone}
\end{equation}
we obtain:
\begin{eqnarray}
K_{\underline{ab}} &=& -\ft 14 \,
R^{\underline{mn}}{}_{\underline{ab}} \, \Gamma_{\underline{mn}} \, +
\,
\mathcal{D}_{[\underline{a}} \, F_{\underline{b}]} \, + \, \ft 12 \, \left[F_{\underline{a}} \, ,
\, F_{\underline{b}} \right]
\label{onespinoRho}
\end{eqnarray}
The tensor-matrix $K_{\underline{ab}}$ is of key importance in the
discussion of compactifications. If it vanishes on a given background
it means that the gravitino field strength can be consistently put to
zero to all orders in $\theta$.s and on its turn this implies that
the $4$--field strength can be chosen constant to all orders in
$\theta$.s This is the case of maximal unbroken supersymmetry. In
this case all curvature components of the Free Differential Algebra
can be chosen constant and we have a superspace whose geometry is
purely described by Maurer Cartan forms of some super coset.
\par
On the other hand if $K_{\underline{ab}}$ does not vanish this
implies that both $\rho_{\underline{ab}}$ and $F_{\underline{abcd}}$
have some non trivial $\theta$-dependence and cannot be chosen
constant. In this case the geometry of superspace is not described by
simple Maurer Cartan forms of some supercoset, since the curvatures
of the FDA are not pure constants. This is the case of fully or
partially broken SUSY and it is the case we want to explore. In the
the $\mathrm{AdS}_4 \times \left( \mathrm{G/H}\right)_7$
compactifications it will turn out that the matrix $K_{ab}$ is
related to the holonomy tensor of the internal manifold $\left(
\mathrm{G/H}\right)_7$.
\par
Let us finally work out the spinor derivative of the Riemann tensor.
Defining:
\begin{equation}
  \mathcal{D} \, R^{\underline{ab}}{}_{\underline{mn}} \, = \,
  \mathcal{D}_{\underline{p}} \,
  R^{\underline{ab}}{}_{\underline{mn}}\, V^{\underline{p}} \, + \,
  \overline{\Psi} \, \Lambda^{\underline{ab}}{}_{\underline{mn}} \,
\label{defiDR}
\end{equation}
from the first of eq.s (\ref{Bianchis}) we obtain:
\begin{equation}
 \Lambda^{\underline{ab}}{}_{\underline{mn}} \, = \,
 \left( \mathcal{D}_{[\underline{m}}\, - \, \overline{F}_{[\underline{m}} \, \right)
  \Theta_{\underline{n}]}^{\phantom{\underline{n}]} \, | \, \underline{ab}} \, + \, 2 \,
  S^{\underline{ab}}\, \rho_{\underline{mn}}
\label{Lambdone}
\end{equation}
where we have introduced the notation:
\begin{eqnarray}
\Theta^{\underline{n}\, | \, \underline{ab}} & = & C \, \left( \bar{\Theta}^{\underline{n}\, | \,
\underline{ab}}\right)^T \, = \, {\rm i} \, \Gamma_{\underline{c}} \, \rho_{\underline{ab}}
\, - \, 2\, {\rm i} \, \Gamma_{[\underline{a}} \, \rho_{\underline{b}] \underline{c}}
 \nonumber\\
\overline{F}_{\underline{a}} & = & C \, \left( F_a\right) ^T \,
C^{-1} \, = \, \ft{\rm i}{24}
\left(\Gamma^{\underline{b_1b_2b_3b_4}}{}_{\underline{a}} \, - \, 8
\,
\delta_{\underline{a}}{}^{[\underline{b_1}}
\Gamma^{\underline{b_2b_3b_4}]}\right)\, F_{\underline{b_1b_2b_3b_4}}
\label{barrati}
\end{eqnarray}
The matrix $K_{\underline{ab}}$ and the spinor
$\Lambda^{\underline{ab}}{}_{\underline{mn}}$ are the crucial objects
we are supposed to compute in each compactification background.
\section{Compactifications of $M$-theory on $\mathrm{{AdS_4}} \times {\mathcal{M}_7}$ backgrounds}
\label{compareview}
We are interested  in compactified
backgrounds where the $11$-dimensional bosonic manifold is of the form:
\begin{equation}
  \mathcal{M}_{11} = \mathcal{M}_4 \, \times \, \mathcal{M}_7
\label{M11split}
\end{equation}
$\mathcal{M}_4$ denoting a four-dimensional maximally symmetric
manifold whose coordinates we denote $x^\mu$ and $\mathcal{M}_7$ a
$7$--dimensional compact manifold whose parameters we denote $y^I$.
Furthermore we assume that in any configuration of the compactified
theory the eleven dimensional vielbein is  split as follows:
\begin{equation}
  V^{\underline{ a}} =\left \{\begin{array}{rclcrcl}
    V^r & = & E^r (x) & ; & r & = & 0,1,2,3 \\
    V^\alpha & = & \Phi^\alpha_{\phantom{I}\beta}(x) \, \left( e^\beta +W^\beta(x)\right)  & ;
    & \alpha,\beta & = & 4,5,6,7,8,9,10 \
  \end{array} \right.
\label{conv1}
\end{equation}
where $E^r(x)$ is a purely $x$--dependent $4$--dimensional
vielbein, $W^\alpha(x)$ is an $x$--dependent $1$--form on $x$-space
describing the Kaluza Klein vectors and the purely $x$--dependent
$7 \times 7$ matrix $\Phi^\alpha_{\phantom{I}\beta}(x)$ encodes part of the
scalar fields of the compactified theory, namely the internal
metric moduli. From these assumptions it
follows that the bosonic field strength is expanded as follows:
\begin{eqnarray}
{ {\mathbf{F}}}^{[4]}_{(Bosonic)} & \equiv & F^{[4]}(x) \,
 + \, F^{[3]}_\alpha(x) \,\wedge \,  V^\alpha + \, F^{[2]}_{\alpha\beta}(x) \,\wedge \,  V^\alpha \,\wedge
\,V^\beta \nonumber\\
 && + \, F^{[1]}_{\alpha\beta\gamma}(x) \,\wedge \,  V^\alpha \,\wedge \,V^\beta \, \wedge \,
 V^\gamma \, + \,  F^{[0]}_{\alpha\beta\gamma\delta}(x) \,\wedge \,  V^\alpha \,\wedge \,V^\beta \, \wedge \,
 V^\gamma \, \wedge \, V^\delta
\label{fb1}
\end{eqnarray}
where $F^{[p]}_{\alpha_{1}\dots \alpha_{4-p}}(x)$ are $x$-space $p$--forms
depending only on $x$.
\par
In bosonic backgrounds with a space--time geometry of the form (\ref{M11split}),
the family of configurations (\ref{conv1}) must satisfy the condition
that by choosing:
\begin{eqnarray}
E^r & = & \mbox{vielbein of a maximally symmetric $4$-dimensional
space time
} \label{maxsym}\\
\Phi^I_{\phantom{I}J}(x) & = & \delta^{I}_{\phantom{I}J} \label{delta=Phi}\\
W^I & = & 0 \label{noKK}\\
F^{[3]}_I(x) & = & F^{[2]}_{IJ}(x) \, = \, F^{[1]}_{IJK}(x) \, = \,
0 \label{viaperlorentz}\\
 F^{[4]}(x) & = & e \, \epsilon_{rstu} \, E^r \, \wedge \, E^s \,
 \wedge \, E^t \, \wedge \, E^u  \quad ; \quad (e = \mbox{constant parameter})\label{freundrub}\\
F^{[0]}_{\alpha\beta\gamma\delta}(x) & = & g_{\alpha\beta\gamma\delta} \, = \, \mbox{constant tensor}
\label{intoflux}
\end{eqnarray}
we obtain an exact \textit{bona fide} solution of the
eleven--dimensional field equations of M-theory.
\par
There are three possible 4--dimensional maximally symmetric Lorentzian
manifolds
\begin{equation}
  \mathcal{M}_4 \, = \, \left\{ \begin{array}{cc}
    \mathcal{M}_4 & \mbox{Minkowsky space} \\
    \mathrm{dS_4} & \mbox{de Sitter space} \\
    \mathrm{AdS_4} & \mbox{anti de Sitter space} \
  \end{array} \right.
\label{threepossib}
\end{equation}
In any case Lorentz invariance imposes
eqs.(\ref{delta=Phi},\ref{noKK},\ref{viaperlorentz}) while
translation invariance imposes that the vacuum expectation value
of the scalar fields $\Phi^\alpha_{\phantom{I}\beta}(x)$ should be a
constant matrix
\begin{equation}
 < \Phi^\alpha_{\phantom{I}\beta}(x) > = \mathcal{A}^{\alpha}_{\beta}
\label{Amatra}
\end{equation}
We are interested in $7$-manifolds that preserve some residual
supersymmetry in $D=4$. This relates to the holonomy of $M_7$ which
has to be restricted in order to allow for the existence of Killing
spinors. In the next subsection we summarize those basic results from Kaluza
Klein literature that are needed in our successive elaborations.
\subsection{M-theory field equations and $7$-manifolds of weak $\mathrm{G_2}$
holonomy \textit{i.e.} Englert $7$-manifolds}
\label{englertspace}
\par
In order to admit at least one Killing spinor or more, the
$7$-manifold $\mathcal{M}_7$ necessarily must have a (weak) holonomy
smaller than $\mathrm{SO(7)}$: at most $\mathrm{G_2}$. The
qualification weak refers to the definition of holonomy appropriate
to compactifications on $\mathrm{AdS_4} \times \mathcal{M}_7$ while the
standard definition of holonomy is appropriate to compactifications
on Ricci flat backgrounds $\mathrm{Mink_4} \times \mathcal{M}_7$. To
explain these concepts that were discovered in the eighties in
contemporary language we have to recall the notion of $G$-structures.
Indeed in the recent literature about flux compactifications the key geometrical notion
exploited by most authors is precisely that of $\mathrm{G}$-structures 
\cite{{tommasiello}}.
\par
Following, for instance, the presentation of \cite{tommasiello},
if $\mathcal{M}_n$ is a differentiable manifold of dimension $n$,
$T\mathcal{M}_n \stackrel{\pi}{\rightarrow} \mathcal{M}_n $ its
tangent bundle and $F\mathcal{M}_n \stackrel{\pi}{\rightarrow}
\mathcal{M}_n $ its frame bundle, we say that $\mathcal{M}_n$
admits a $\mathrm{G}$-structure when the structural group of
$F\mathcal{M}_n$ is reduced from the generic
$\mathrm{GL(n,\mathbb{R})}$ to a proper subgroup $\mathrm{G}
\subset \mathrm{GL(n,\mathbb{R})}$. Generically, tensors on
$\mathcal{M}_n$ transform in representations of the structural
group $\mathrm{GL(n,\mathbb{R})}$. If a $\mathrm{G}$-structure
reduces this latter to $\mathrm{G} \subset
\mathrm{GL(n,\mathbb{R})}$, then the decomposition of an
 irreducible representation of $\mathrm{GL(n,\mathbb{R})}$, pertaining
to a certain tensor $t^{p}$, with respect to the subgroup
$\mathrm{G}$ may contain singlets. This means that on such a
manifold $\mathcal{M}_n$ there may exist a certain tensor $t^{p}$
which is $\mathrm{G}$--invariant, and therefore globally defined.
As recalled in \cite{tommasiello} existence of a Riemannian metric
$g$ on $\mathcal{M}_n$ is equivalent to a reduction of the
structural group $\mathrm{GL(n,\mathbb{R})}$ to $\mathrm{O(n)}$,
namely to an $\mathrm{O(n)}$-structure. Indeed, one can reduce the
frame bundle by introducing orthonormal frames, the vielbein
$e^I$, and, written in these frames, the metric is the
$\mathrm{O(n)}$ invariant tensor $\delta_{IJ}$.  Similarly
orientability corresponds to an $\mathrm{SO(n)}$-structure and the
existence of spinors on spin manifolds corresponds to a
$\mathrm{Spin}(n)$-structure.
\par
In the case of seven dimensions, an orientable Riemannian manifold
$\mathcal{M}_7$, whose frame bundle has generically an
$\mathrm{SO(7)}$ structural group admits a
$\mathrm{G_2}$-structure if and only if, in the basis provided by
the orthonormal frames $\mathcal{B}^\alpha$, there exists an  antisymmetric
$3$-tensor $\phi_{\alpha\beta\gamma\delta}$ satisfying the algebra of the octonionic
structure constants:
\begin{eqnarray}
  \phi_{\alpha\beta\kappa} \, \phi_{\gamma\delta\kappa} & = & \ft {1}{18} \, \delta^{\gamma\delta}_{\alpha\beta} \,
  - \, \ft {2}{3} \, \phi^{\star}_{\alpha\beta\gamma\delta} \nonumber\\
  - \ft {1}{6} \, \epsilon_{\kappa\rho\sigma\alpha\beta\gamma\delta} \, \phi^{\star}_{\alpha\beta\gamma\delta} & = & \phi_{\kappa\rho\sigma}
\label{octonio}
\end{eqnarray}
which is invariant, namely it is the same in all local
trivializations of the $\mathrm{SO(7)}$ frame bundle.
This corresponds to the algebraic definition of
$\mathrm{G_2}$ as that subgroup of $\mathrm{SO(7)}$ which acts as
an automorphism group of the octonion algebra.
Alternatively $\mathrm{G_2}$ can be defined as the stability subgroup of the
$8$-dimensional spinor representation of $\mathrm{SO(7)}$.  Hence
we can equivalently state that a manifold $\mathcal{M}_7$ has a
$\mathrm{G_2}$-structure if there exists at least an invariant
spinor $\eta$, which is the same in all local trivializations of the
$\mathrm{Spin(7)}$ spinor bundle.
\par
In terms of
this invariant  spinor the invariant 3--tensor $\phi_{\rho\sigma\kappa}$ has the form:
\begin{equation}
  \phi^{\rho\sigma\kappa} = \ft 16 \, \eta^T \, \tau^{\rho\sigma\kappa} \, \eta
\label{etaTeta}
\end{equation}
and eq.(\ref{etaTeta}) provides the relation between the two
definitions of the $\mathrm{G_2}$-structure.
\par
On the other hand the manifold has not only a $\mathrm{G_2}$--structure, but also  $\mathrm{G_2}$--holonomy
if the  invariant three--tensor $\phi_{\alpha\beta\kappa}$ is covariantly constant. Namely we must have:
\begin{eqnarray}
0 & = & \nabla \phi^{\alpha\beta\gamma} \, \equiv \, d\phi^{\alpha\beta\gamma} \, + \, 3 \, \mathcal{B}^{\kappa[\alpha} \,  \phi^{\beta\gamma]\kappa}
\label{octonioHOLO}
\end{eqnarray}
where the $1$-form $\mathcal{B}^{\alpha\beta}$ is the spin connection of $\mathcal{M}_7$.
Alternatively the manifold has $\mathrm{G_2}$--holonomy if the
invariant spinor $\eta$ is covariantly constant, namely if:
\begin{equation}
  \exists \, \eta \, \in \, \Gamma(\mathrm{Spin} \mathcal{M}_7 ,
  \mathcal{M}_7) \quad \backslash \quad  0 \, = \,\nabla \, \eta \, \equiv d\eta
  -\ft 14 \, \mathcal{B}^{\alpha\beta} \, \tau_{\alpha\beta} \, \eta
\label{covspinor}
\end{equation}
where $\tau^\alpha$ ($\alpha =1,\dots, 7$) are the $8 \times 8$ gamma matrices of the $\mathrm{SO(7)}$
Clifford algebra.
The relation between the two definitions (\ref{octonioHOLO}) and
(\ref{covspinor}) of  $\mathrm{G_2}$-holonomy  is the same as for the
two definitions of the $\mathrm{G_2}$-structure, namely it is given by eq.(\ref{etaTeta}).
As a consequence of its own definition a Riemannian $7$-manifold with
$\mathrm{G_2}$ holonomy  is Ricci flat. Indeed the integrability condition of eq.(\ref{covspinor})
yields:
\begin{equation}
  \mathcal{R}^{\alpha\beta}_{\phantom{\alpha\beta}\gamma\delta} \, \tau_{\alpha\beta} \, \eta \, = \, 0
\label{holonomia1}
\end{equation}
where $\mathcal{R}^{\alpha\beta}_{\phantom{\alpha\beta}\gamma\delta}\,$ is the Riemann tensor of
$\mathcal{M}_7$. From eq.(\ref{holonomia1}), by means of a few simple
algebraic manipulations one obtains two results:
\begin{itemize}
  \item{ The curvature $2$-form
\begin{equation}
  \mathcal{R}^{\alpha\beta}\, \equiv \, \mathcal{R}^{\alpha\beta}_{\phantom{\alpha\beta}\gamma\delta} \, \mathcal{B}^\gamma \, \wedge
  \, \mathcal{B}^\delta
\label{curvadue}
\end{equation}
is $\mathrm{G_2}$ Lie algebra valued,  namely it satisfies the condition:
\begin{equation}
  \phi^{\kappa\alpha\beta} \, \mathcal{R}^{\alpha\beta} \, = \, 0
\label{G2condition}
\end{equation}
which projects out the $\mathbf{7}$ of $\mathrm{G_2}$ from the $\mathbf{21}$
of $\mathrm{SO(7)}$ and leaves with the adjoint $\mathbf{14}$.}
  \item {The internal Ricci tensor is zero:
\begin{equation}
  \mathcal{R}^{\alpha\kappa}_{\phantom{\alpha\kappa}\beta\kappa} \, =\, 0
\label{riccipiatto}
\end{equation}
   }
\end{itemize}
Next we consider the bosonic field equations of $M$-theory, namely
the first and the last of eq.s ( \ref{fieldeque} ). We
make the compactification ansatz (\ref{M11split}) where $\mathcal{M}_4$ is one of the three possibilities mentioned
in eq.(\ref{threepossib}) and all of
eq.s(\ref{delta=Phi}-\ref{intoflux}) hold true.  Then we split the
rigid index range as follows:
\begin{equation}
  {\underline{a}},{\underline{b}},{\underline{c}},\dots = \cases{
  \alpha,\beta,\gamma,\dots \, \, = 4,5,6,7,8,9,10  \, = \mbox{$\mathcal{M}_7$ indices}\cr
  r,s,t,\dots  \quad = 0,1,2,3 \quad\quad =
  \mbox{$\mathcal{M}_4$ indices} \cr}
\label{indexsplit}
\end{equation}
and by following the conventions employed in \cite{su3su2u1} and using the
results obtained in the same paper, we conclude that  the compactification ansatz
reduces the system of the first and last of (\ref{fieldeque})
to the following one:
\begin{eqnarray}
{R}^{rs}_{\phantom{ab}tu} & = & \lambda \, \delta^{rs}_{tu} \label{AdScurva}\\
\mathcal{R}^{\alpha\kappa}_{\phantom{\alpha\kappa}\beta\kappa}  & = & 3 \, \nu \, \delta^{\alpha}_{\beta} \label{M7curva}\\
F_{rstu} & = & e \, \epsilon_{rstu} \label{externalflux}\\
g_{\alpha\beta\gamma\delta} & = &f\, \mathcal{F}_{\alpha\beta\gamma\delta}  \label{internalflux}\\
\mathcal{F}^{\alpha\kappa\rho\sigma} \, \mathcal{F}_{\beta\kappa\rho\sigma} & = & \mu \, \delta^{\alpha}_{\beta} \label{gsquare}\\
\mathcal{D}^\mu \, \mathcal{F}_{\mu\kappa\rho\sigma} & = & \ft 12 \, e \, \epsilon_{\kappa\rho\sigma\alpha\beta\gamma\delta} \,
\mathcal{F}^{\alpha\beta\gamma\delta}\label{englert}
\end{eqnarray}
 Eq. (\ref{M7curva}) states that the internal manifold
$\mathcal{M}_{7}$ must be an Einstein space. Eq.s
(\ref{externalflux}) and (\ref{internalflux}) state that there is
a flux of the four--form both on $4$--dimensional space-time
$\mathcal{M}_4$ and on the internal manifold $\mathcal{M}_7$. The
parameter $e$, which fixes the size of the flux on the
four--dimensional space and was already introduced in
eq.(\ref{freundrub}), is called the Freund-Rubin parameter
\cite{freundrubin}. As we are going to show, in the case that a
non vanishing $\mathcal{F}^{\alpha\beta\gamma\delta}$ is required to exist, eq.s
(\ref{gsquare}) and (\ref{englert}),  are equivalent to the
assertion that the manifold $\mathcal{M}_7$ has weak
$\mathrm{G_2}$ holonomy rather than $\mathrm{G_2}$--holonomy, to
state it in modern parlance \cite{weakG2}. In paper
\cite{fermionMassSpectrum}, manifolds admitting such a structure
were instead named \textit{Englert spaces} and the underlying
notion of weak $\mathrm{G_2}$ holonomy was already introduced
there with the different name of \textit{ de Sitter
$\mathrm{SO(7)}^+$ holonomy}.
\par
Indeed eq.(\ref{englert}) which, in the language of the early
eighties was named Englert equation \cite{englertsolution} and
which is nothing else but the first of equations (\ref{fieldeque}), upon
substitution of the Freund Rubin ansatz (\ref{externalflux}) for
the external flux, can be recast in the following more revealing
form: Let
\begin{equation}
  \Phi^\star \, \equiv \, \mathcal{F}_{\alpha\beta\gamma\delta} \, \mathcal{B}^\alpha \, \wedge \,
  \mathcal{B}^\beta
  \, \wedge \, \mathcal{B}^\gamma  \, \wedge \, \mathcal{B}^\delta
\label{phistarra}
\end{equation}
be a the constant $4$--form on $\mathcal{M}_7$ defined by our non
vanishing flux, and let
\begin{equation}
  \Phi \, \equiv \,\ft {1}{24} \epsilon_{\alpha\beta\gamma\kappa\rho\sigma\tau} \, \mathcal{F}_{\kappa\rho\sigma\tau} \, \mathcal{B}^\alpha \, \wedge \,
  \mathcal{B}^\beta
  \, \wedge \, \mathcal{B}^\gamma  \,
\label{nophistarra}
\end{equation}
be its dual. Englert eq.(\ref{englert}) is just the same as writing:
\begin{eqnarray}
  d \Phi & = & 12\, e \, \Phi^\star \nonumber\\
   d \Phi^\star & = & 0
\label{ollaolla}
\end{eqnarray}
When the Freund Rubin parameter vanishes $e=0$ we recognize in
eq.(\ref{ollaolla}) the statement that our internal manifold
$\mathcal{M}_7$ has $\mathrm{G_2}$-holonomy and hence  it is Ricci
flat.  Indeed $\Phi$ is the $\mathrm{G_2}$ invariant and
covariantly constant form defining $\mathrm{G_2}$-structure and
$\mathrm{G_2}$-holonomy. On the other hand the case $e\ne 0$
corresponds to the  weak $\mathrm{G_2}$ holonomy. Just as we
reduced the existence  of a closed three-form $\Phi$ to the
existence of a $\mathrm{G_2}$ covariantly constant spinor
satisfying eq.(\ref{covspinor}) which allows to set the
identification (\ref{etaTeta}), in the same way eq.s
(\ref{ollaolla}) can be solved \textit{if and only if} on
$\mathcal{M}_7$ there exist a weak Killing spinor $\eta$
satisfying the following defining condition:
\begin{eqnarray}
  \mathcal{D}_\alpha \, \eta & = & m \, e \, \tau_\alpha \, \eta
\label{weakkispinor}\\
\null &\Updownarrow & \null \nonumber\\
D\eta\equiv(d - {1\over 4} \mathcal{B}^{\alpha\beta}\tau_{\alpha\beta})\eta
&=&
m \, e\, \mathcal{B}^\alpha \tau_\alpha\eta \label{kilspineq7dim}
\end{eqnarray}
 where $m$ is a numerical constant
and $e$ is the Freund-Rubin parameter, namely the only scale which
at the end of the day will occur in the solution.\par The
integrability of the above equation implies that the Ricci tensor
be proportional to the identity, namely that the manifold is an
Einstein manifold and furthermore fixes the proportionality
constant:
\begin{equation}
  \mathcal{R}^{\alpha\kappa}_{\phantom{IM}\beta\kappa} = 12 \, m^2 \, e^2 \, \delta^{\alpha}_{\beta} \quad
  \longrightarrow \, \nu = \, 12 \, m^2 \, e^2
\label{fixingnu}
\end{equation}
In case such a spinor exists, by setting:
\begin{equation}
  g_{\alpha\beta\gamma\delta} = \mathcal{F}_{\alpha\beta\gamma\delta} = \eta^T \, \tau_{\alpha\beta\gamma\delta} \eta \, = \, 24 \,
  \phi^\star_{\alpha\beta\gamma\delta}
\label{agnisco1}
\end{equation}
we find that Englert equation (\ref{englert}) is satisfied, provided
we have:
\begin{equation}
  m = - \frac{3}{2}
\label{tremezzi}
\end{equation}
In this way Maxwell equation, namely the first of
(\ref{fieldeque}) is solved.
Let us also note, as the authors of \cite{fermionMassSpectrum} did
many years ago, that condition (\ref{weakkispinor}) can also be
interpreted in the following way. The spin-connection $\mathcal{B}^{\alpha\beta}$
plus the vielbein $\mathcal{B}^\gamma$ define on any non Ricci flat $7$-manifold  $\mathcal{M}_7$ a
connection which is actually $\mathrm{SO(8)}$ rather than $\mathrm{SO(7)}$ Lie algebra
valued. In other words we have a principal $\mathrm{SO(8)}$ bundle
which leads to an $\mathrm{SO(8)}$ spin bundle of which $\eta$ is a
covariantly constant section:
\begin{equation}
 0 \, = \, \nabla^{\mathrm{SO(8)}}\eta =  \Big(\nabla^{\mathrm{SO(7)}} \,-\,  m \, e \, \mathcal{B}^\alpha\,  \tau_\alpha\Big) \,
  \eta
\label{so8cov}
\end{equation}
The  existence of $\eta$  implies a reduction of the $\mathrm{SO(8)}$-bundle.
Indeed the stability subgroup of an $\mathrm{SO(8)}$ spinor is a well
known subgroup $\mathrm{SO(7)}^+$ different from the standard $\mathrm{SO(7)}$ which,
instead, stabilizes the vector representation. Hence the so named weak $\mathrm{G_2}$
holonomy of the $\mathrm{SO(7)}$ spin connection $\mathcal{B}^{\alpha\beta}$ is the same
thing as the $\mathrm{SO(7)}^+$ holonomy of the  $\mathrm{SO(8)}$ Lie algebra
valued \textit{de Sitter connection} $\left\{\mathcal{B}^{\alpha\beta},
\mathcal{B}^\gamma\right\}$ introduced in \cite{fermionMassSpectrum} and normally
discussed in the old literature on Kaluza Klein Supergravity.
\par
We have solved Maxwell equation, but we still have to  solve
Einstein equation, namely the last of (\ref{fieldeque}). To this effect we
note that:
\begin{equation}
  \mathcal{F}_{\beta\kappa\rho\sigma}\,\mathcal{F}^{\alpha\kappa\rho\sigma} \, = \, 24 \,
  \delta^\alpha_\beta
  \quad \Longrightarrow \quad \mu = 24
\label{mufixed}
\end{equation}
and we observe that Einstein equation reduces to the following two
conditions on the parameters (see \cite{su3su2u1} for details) :
\begin{eqnarray}
\ft 32 \, \lambda & = & - \left( 24 \, e^2 + \ft 7 2 \, \mu \, f^2 \right)  \nonumber\\
3 \, \nu  & = & 12 \, e^2 + \ft 52 \, \mu \, f^2
\label{twocondos}
\end{eqnarray}
>From eq.s (\ref{twocondos}) we conclude that there are only three
possible kind of solutions.
\begin{description}
  \item[a] The flat solutions of type
\begin{equation}
  \mathcal{M}_{11} \, = \, \mathrm{Mink_4} \, \otimes \, \underbrace{\mathcal{M}_7}_{\mbox{Ricci flat}}
\label{minktimespiat}
\end{equation}
where both $D=4$ space-time and the internal $7$-space are Ricci
flat. These compactifications correspond to  $e=0$ and $F_{\alpha\beta\gamma\delta} =
0 \, \Rightarrow \, g_{\alpha\beta\gamma\delta} = 0$.
  \item[b] The Freund Rubin solutions of type
\begin{equation}
  \mathcal{M}_{11} \, = \, \mathrm{AdS}_4 \, \otimes \, \underbrace{\mathcal{M}_7}_{\mbox{Einst. manif.}}
\label{adstimesEinst}
\end{equation}
These correspond to anti de Sitter space in $4$-dimensions, whose
radius is fixed by the Freund Rubin parameter $e \ne 0 $ times any
Einstein manifold in $7$--dimensions with no internal flux, namely
$g_{\alpha\beta\gamma\delta} = 0$.
In this case from eq.(\ref{twocondos})  we uniquely
obtain:\begin{eqnarray}
R^{rs}_{\phantom{ab}tu} & = & -16 \, e^2 \, \delta^{rs}_{tu} \label{AdScurvaFRUB}\\
\mathcal{R}^{\alpha\kappa}_{\beta\kappa} & = & 12 \, e^2 \,  \delta^{\alpha}_{\beta} \label{M7curvaFRUB}\\
F_{rstu} & = & e \, \epsilon_{rstu} \label{external fluxFRUB}\\
F_{\alpha\beta\gamma\delta} & = & 0 \label{internalfluxFRUB}
\end{eqnarray}
  \item[c] The Englert type solutions
  \begin{equation}
  \mathcal{M}_{11} \, = \, \mathrm{AdS}_4 \, \otimes \,
  \underbrace{\mathcal{M}_7}_{\begin{array}{c}
    \mbox{Einst. manif.} \\
    \mbox{weak $\mathrm{G_2}$ hol} \
  \end{array}}
\label{adstimeG2}
\end{equation}
These correspond to anti de Sitter space in $4$-dimensions ($e \ne
0$) times a $7$--dimensional Einstein manifold which is
necessarily of weak $\mathrm{G_2}$ holonomy in order to support a
consistent non vanishing internal flux $g_{\alpha\beta\gamma\delta}$.
In this case combining eq.s (\ref{twocondos}) with the previous ones we uniquely
obtain:
\begin{equation}
  \lambda = -30 \, e^2 \quad ; \quad f = \pm \ft 12 \, e
\label{soluzia}
\end{equation}
\end{description}
\par
As we already mentioned in the introduction there exist several
compact manifolds of weak $\mathrm{G_2}$ holonomy. In particular
all the coset manifolds $\mathcal{G}/\mathcal{H}$ of weak
$\mathrm{G_2}$ holonomy were classified and studied in the Kaluza
Klein supergravity age
\cite{Awada:1982pk,su3su2u1,D'Auria:1983vy,Castellani:1983tc,classificoleo,frenico,multanna,fermionMassSpectrum,
roundspectrum,roundseven}
and they were extensively reconsidered in the context of the
AdS/CFT correspondence
\cite{Billo:2000zs,Billo:2000zr,Fre':1999xp,Fabbri:1999hw,Fabbri:1999mk}.
\par
In the present paper we study the supergauge completion of
compactifications of the Freund Rubin type, namely on eleven-manifolds of the form:
\begin{equation}
  \mathcal{M}_{11} = \mathrm{AdS_4} \, \times \,
  \frac{\mathcal{G}}{\mathcal{H}}
\label{M11splittus}
\end{equation}
with no internal flux $g_{\alpha\beta\gamma\delta}$ switched on.
As it was extensively explained in \cite{universal} and further
developed in
\cite{Billo:2000zs,Billo:2000zr,Fre':1999xp,Fabbri:1999hw,Fabbri:1999mk},
if the compact coset $\mathcal{G}/\mathcal{H}$ admits $\mathcal{N} \le 8$
Killing spinors $\eta_A$, namely $N \le 8$ independent solutions of
equation (\ref{weakkispinor}) with $m=1$, then the isometry group $\mathcal{G}$
is necessarily of the form:
\begin{equation}
  \mathcal{G} = \mathrm{SO(\mathcal{N})} \, \times \, \mathrm{G}_{flavor}
\label{Sgruppo}
\end{equation}
where $\mathrm{G}_{flavor}$ is some appropriate Lie group. In this
case the isometry supergroup of the considered M-theory background is:
\begin{equation}
  \mathrm{Osp(\mathcal{N} \, | \, 4 }) \, \times \, \mathrm{G}_{flavor}
\label{oppala}
\end{equation}
and the spectrum of fluctuations of the background arranges into $\mathrm{Osp(\mathcal{N} \, | \, 4
})$ supermultiplets furthermore assigned to suitable representations
of the bosonic flavor group.
\section{The $\mathrm{SO(8)}$ spinor bundle and the holonomy tensor}
We come next to discuss a very important property of $7$--manifolds
with a spin structure which plays a crucial role in understanding
the supergauge completion. This is the existence of an $\mathrm{SO(8)}$
vector bundle whose non trivial connection is defined by the
riemannian structure of the manifold. To introduce this point and in
order to illustrate its relevance to our problem we begin by
considering a basis of $D=11$ gamma matrices well adapted to the
compactification on $\mathrm{AdS_4} \times \mathcal{M}_7$.
\subsection{The well adapted basis of gamma matrices}
According to the tensor product representation well adapted to the
compactification, the $D=11$ gamma matrices can be written as follows:
\begin{eqnarray}
\Gamma_a & = & {\gamma}_a \, \otimes \, \mathbf{1}_{8 \times 8} \quad (a=0,1,2,3) \nonumber\\
\Gamma_{3+\alpha} & = & {\gamma}_5 \, \otimes \, \tau_\alpha \quad (\alpha=1,\dots,7)
\label{gammone}
\end{eqnarray}
where, following \cite{antonpietro} and the old Kaluza Klein
supergravity literature
\cite{fermionMassSpectrum,universal,Castellani:1983tc} the matrices
$\tau_\alpha$ are the real antisymmetric realization of the $\mathrm{SO(7)}$
Clifford algebra with negative metric:
\begin{equation}
  \left\{ \tau_\alpha \, , \, \tau_\beta \right\} \, = - \, 2 \, \delta_{\alpha\beta}
  \, \quad ; \quad \tau_\alpha \, = \, - \left( \tau_\alpha \right)^T
\label{taupiccole}
\end{equation}
In this basis the charge conjugation matrix is given by:
\begin{equation}
  C \, = \, \mathcal{C} \, \otimes \,\mathbf{1}_{8 \times 8}
\label{fertilissima}
\end{equation} where $\mathcal{C}$ is the charge conjugation matrix in
$d=4$:
\begin{equation}
  \mathcal{C} \, \gamma_a \, \mathcal{C} ^{-1} \, = \, - \gamma_a^T
  \quad ; \quad \mathcal{C} ^T = - \mathcal{C}
\label{Cstorta}
\end{equation}
\subsection{The $\so(8)$-connection and the holonomy
tensor}
Next we observe that using these matrices the covariant derivative
introduced in equation (\ref{so8cov}) defines a universal
$\so(8)$-connection on the spinor bundle which is given once the
riemannian structure, namely the vielbein and the spin connection are
given $\left\{ \mathcal{B}^\alpha , B^{\alpha \beta }\right\} $:
\begin{equation}
  \mathbf{U}^{\so(8)} \, \equiv \, -\ft 14 \, \mathcal{B}^{\alpha \beta } \,
  \tau_{\alpha \beta } \, - \, e \, \mathcal{B}^\alpha \, \tau_\alpha
\label{so8Ucon1}
\end{equation}
More precisely and following the index conventions presented in
appendix \ref{indexconven}, let $\zeta_{\underline{A}}$ be an
orthonormal basis:
\begin{equation}
  \overline{\zeta}_{{A}} \, \zeta_{{B}} \, = \,
  \delta_{{AB}}
\label{ortobaseseczia}
\end{equation}
of sections of the spinor bundle over the Einstein manifold
$\mathrm{M}_7$. Any spinor can be written as a linear combination of
these sections that are real. Furthermore the bar operation in this
case is simply the transposition.  Hence, if we consider the $\so(8)$
covariant derivative of any of these sections, this is a spinor and, as such,
it can be expressed as a linear combinations of the same:
\begin{equation}
  \nabla^{\so(8)}\, \zeta_{{A}} \, \equiv \, \left(
  d + \mathbf{U}^{\so(8)} \right)  \zeta_{{A}} \, = \,
  \mathbf{U}_{{AB}} \, \zeta_{\mathrm{{B}}}
\label{Ugrosso}
\end{equation}
According to standard lore the $1$-form valued, antisymmetric $8 \times 8$
matrix $\mathbf{U}_{{AB}}$ defined by eq.(\ref{Ugrosso}) is the
$\so(8)$-connection in the chosen basis of sections. If the manifold
$\mathcal{M}_7$ admits $\mathcal{N}$ Killing spinors, then it follows
that we can choose an orthonormal basis where the first $\mathcal{N}$ sections
are Killing spinors:
\begin{equation}
  \zeta_{\underline{A}} = \eta_{\underline{A}} \quad ; \, \nabla^{\so(8)}\, \eta_{{\underline{A}}} \,
  = \, 0 \quad , \quad \underline{A}=1, \dots \, , \, \mathcal{N}
\label{deta=0}
\end{equation}
and the remaining $8-\mathcal{N}$ elements of the basis, whose covariant derivative
does not vanish are orthogonal to the Killing spinors:
\begin{eqnarray}
  \zeta_\Lambda & = & \xi_{\overline{A}} \quad ; \, \nabla^{\so(8)}\, \xi_{{\overline{A}}} \,
  \ne \, 0 \quad , \quad \overline{A}=1, \dots \, , \, 8 -
  \mathcal{N}\nonumber\\
 \overline{ \xi}_{\overline{B}} \, \eta_{\underline{A}} & = & 0 \nonumber\\
  \overline{\xi}_{\overline{B}} \, \xi_{\overline{C} }& = & \delta_{\overline{BC}}
\label{mezzaquimezzala}
\end{eqnarray}
It is then evident from eq.s (\ref{deta=0}) and
(\ref{mezzaquimezzala}) that the $\so(8)$-connection $
\mathbf{U}_{AB}$ takes values only in a subalgebra $\so(8-\mathcal{N}) \subset
\so(8)$ and has the following block diagonal form:
\begin{equation}
\mathbf{U}_{{AB}} \, = \,  \left( \begin{array}{c|c}
    0 & 0 \\
    \hline
    0 &\mathbf{ U}_{\overline{AB} } \
  \end{array}\right)
\label{Upiccola}
\end{equation}
Squaring the $\mathrm{SO(8)}$-covariant derivative, we find
\begin{eqnarray}
\nabla^2 \, \zeta_{{A}} & = & \underbrace{\left( d\mathbf{U}_{{AB}} \,
- \, \mathbf{U}_{\underline{AC}} \,
\wedge \, \mathbf{U}_{{CB}} \right)}_{\mathcal{F}_{{AB}}[\mathbf{U}]} \,
\zeta_{\underline{B}} \nonumber\\
\null & = & \, - \ft 14 \,  \underbrace{\left( \mathcal{R}^{\gamma\delta}{}_{\alpha\beta} \, - \,
4\, e^2 \, \delta^{\gamma\delta}{}_{\alpha\beta}\right
)}_{\mathcal{C}^{\gamma\delta}{}_{\alpha\beta}} \,
\tau_{\gamma\delta} \, \zeta_{\underline{A}}
\label{fortedeimarmi}
\end{eqnarray}
where $\mathcal{C}^{\gamma\delta}{}_{\alpha\beta}$ is the so called
holonomy tensor, essentially identical with the Weyl tensor of the
considered Einstein $7$-manifold.
\subsection{The holonomy tensor and superspace}
As a further preparation to our subsequent discussion of the gauge completion let us now consider the
form taken on the $\mathrm{AdS_4} \times \mathcal{G}/\mathcal{H} $ backgrounds by the operator $K_{\underline{ab}}$
introduced in equation (\ref{festone})
and governing the mechanism of supersymmetry breaking. We will see
that it is just simply related to the holonomy tensor discussed in
the previous section, namely to the field strength of the
$\mathrm{SO(8)}$-connection on the spinor bundle.
To begin with we calculate the operator $F_{\underline{a}}$
introduced in eq.s (\ref{defHCF},\ref{defT}). Explicitly using the
well adapted basis (\ref{gammone}) for gamma matrices we find:
\begin{equation}
  F_{\underline{a}} \, = \, \cases{F_a \, = \, - 2\, e\, \gamma_a \,
  \gamma_5 \, \otimes \, \mathbf{1}_8 \cr
  F_\alpha \, = \, - e \, \mathbf{1}_4 \, \otimes \, \tau_\alpha \cr}
\label{FaOnads4}
\end{equation}
Using this input we obtain:
\begin{equation}
  K_{\underline{ab}} \, = \, \cases{
K_{ab} \, = \, 0 \cr
K_{a\beta} \, = \, 0 \cr
K_{\alpha \beta } \, = \, -\ft 14 \, \underbrace{\left( \mathcal{R}^{\gamma\delta}{}_{\alpha\beta} \, - \, 4 \, e^2 \,
\delta^{\gamma\delta}{}_{\alpha\beta}\right)}_{C^{\gamma\delta}{}_{\alpha\beta}}
\, \tau_{\gamma\delta} \cr
  }
\label{holotensor}
\end{equation}
Where the tensor $C^{\gamma\delta}{}_{\alpha\beta}$ defined by the
above equation is named the holonomy tensor and it is an intrinsic
geometric property of the compact internal manifold $\mathcal{M}_7$.
As we see the holonomy tensor vanishes only in the case of $\mathcal{M}_7 =
\mathcal{S}^7$ when the Riemann tensor is proportional to an
antisymmetrized Kronecker delta, namely, when the internal Einstein $7$-manifold is
maximally symmetric. The holonomy tensor is a $21 \times 21 $ matrix
which projects the $\mathrm{SO(7)}$ Lie algebra to a subalgebra:
\begin{equation}
  \mathbb{H}_{hol} \, \subset \, \mathrm{SO(7)}
\label{Hhol}
\end{equation}
with
respect to which the $8$-component spinor representation should
contain singlets in order for unbroken supersymmetries to survive.
Indeed the holonomy tensor appears in the integrability condition for
Killing spinors. Indeed squaring the defining equation of Killing
spinors with $m=1$ we get the consistency condition:
\begin{equation}
  C^{\gamma\delta}{}_{\alpha\beta} \, \tau_{\gamma\delta}  \, \eta \,
  \, = \, 0
\label{frullo}
\end{equation}
which states that the Killing spinor directions are in the kernel of
the operators $ C^{\gamma\delta}{}_{\alpha\beta} \,
\tau_{\gamma\delta}$, namely are singlets of the subalgebra $\mathbb{H}_{hol}
$ generated by them.
\par
In view of this we conclude that the gravitino field strength has the
following structure:
\begin{equation}
  \rho_{\underline{ab}} \, = \, \cases{\rho_{ab} \, = \,0
  \cr
  \rho_{a \beta } \, = \, 0 \cr
  \rho_{\alpha \beta }\, \neq \, 0 \, ; \, \cases{\mbox{zero at $\theta=0$} \cr
  \mbox{depends only on the broken $\theta$.s}\cr}\cr }
\label{rhostructa}
\end{equation}
As a preparation for our next coming discussion it is now useful to
remind the reader that the list of homogeneous $7$-manifolds
$\mathcal{G}/\mathcal{H}$ of Englert type which preserve at least two
supersymmetries ($\mathcal{N} \ge 2$) is extremely short. It
consists of the sasakian or tri-sasakian homogeneous manifolds which
are displayed in table \ref{sasakiani}. For these cases our strategy in order to obtain
the supergauge completion will be based on a superextension of the sasakian fibration. The cases with $\mathcal{N}=1$ are
somewhat more involved since such a weapon is not in our stoke. These
cases are also ultra-few and they are displayed in table
\ref{n=1casi}.
\begin{table}
  \centering
  {\small \begin{tabular}{|c||c|c|c|l|}
\hline
 $\mathcal{N}$ & Name & Coset &$\begin{array}{c}
   \mbox{Holon.} \\
   \so(8) \mbox{ bundle } \
 \end{array}$ & Fibration \\
  \hline
  8 & $\mathbb{S}^7$ & $\frac{\mathrm{SO(8)}}{\mathrm{SO(7)}}$ & 1 &
  $ \left \{ \begin{array}{l}
    \mathbb{S}^7 \, \stackrel{\pi}{\Longrightarrow} \, \mathbb{P}^3 \\
    \forall \, p \, \in \, \mathbb{P}^3 \, ; \, \pi^{-1}(p) \, \sim \, \mathbb{S}^1\\
  \end{array}  \right. $ \\
  \hline
  2 & $M^{111}$ & $\frac{\mathrm{SU(3)\times SU(2)\times U(1)}}{\mathrm{SU(2) \times U(1) \times U(1) }}$ & $\mathrm{SU(3)}$ &
  $ \left \{ \begin{array}{l}
    M^{111} \, \stackrel{\pi}{\Longrightarrow} \, \mathbb{P}^2  \, \times \, \mathbb{P}^1\\
    \forall \, p \, \in \, \mathbb{P}^2  \, \times \, \mathbb{P}^1\, ; \, \pi^{-1}(p) \, \sim \, \mathbb{S}^1\\
  \end{array}  \right. $ \\
  \hline
   2 & $Q^{111}$ & $\frac{\mathrm{SU(2)\times SU(2)\times SU(2) \times U(1)}}{\mathrm{U(1) \times U(1) \times U(1) }}$ & $\mathrm{SU(3)}$ &
  $ \left \{ \begin{array}{l}
    Q^{111} \, \stackrel{\pi}{\Longrightarrow} \, \mathbb{P}^1  \, \times \, \mathbb{P}^1\, \times \,\mathbb{P}^1  \\
    \forall \, p \, \in \, \mathbb{P}^1  \, \times \, \mathbb{P}^1\, \times \,\mathbb{P}^1  \, ; \, \pi^{-1}(p) \, \sim \, \mathbb{S}^1\\
  \end{array}  \right. $ \\
  \hline
   2 & $V^{5,2}$ & $\frac{\mathrm{SO(5)}}{\mathrm{SO(2)}}$ & $\mathrm{SU(3)}$ &
  $ \left \{ \begin{array}{l}
    V^{5,2} \, \stackrel{\pi}{\Longrightarrow} \, M_a \, \sim \, \mbox{quadric in } \mathbb{P}^4  \\
    \forall \, p \, \in \, \, M_a \, \, ; \, \pi^{-1}(p) \, \sim \, \mathbb{S}^1\\
  \end{array}  \right. $ \\
  \hline
   3 & $N^{010}$ & $\frac{\mathrm{SU(3)\times SU(2)}}{\mathrm{SU(2)\times U(1)}}$ & $\mathrm{SU(2)}$ &
  $ \left \{ \begin{array}{l}
    N^{010} \, \stackrel{\pi}{\Longrightarrow} \, \mathbb{P}^2  \\
    \forall \, p \, \in \, \, \mathbb{P}^2 \, \, ; \, \pi^{-1}(p) \, \sim \, \mathbb{S}^3\\
  \end{array}  \right. $ \\
  \hline
  \end{tabular}}
  \caption{The homogeneous $7$-manifolds that admit at least $2$ Killing spinors are all sasakian or tri-sasakian. This is evident
  from the fibration structure of the $7$-manifold, which is either a fibration in circles $\mathbb{S}^1$ for the $\mathcal{N}=2$ cases or a
  fibration in $\mathbb{S}^3$ for the unique
  $\mathcal{N}=3$ case corresponding to the $\mathrm{N}^{010}$ manifold}\label{sasakiani}
\end{table}
\vskip 0.2cm
\begin{table}
  \centering
  {\small \begin{tabular}{|c||c|c|c|}
\hline
 $\mathcal{N}$ & Name & Coset &$\begin{array}{c}
   \mbox{Holon.} \\
   \so(8) \mbox{ bundle } \
 \end{array}$ \\
  \hline
   \null & \null & \null & \null \\
  1 & $\mathbb{S}^7_{squashed}$ & $\frac{\mathrm{SO(5)\times SO(3)}}{\mathrm{SO(3)\times SO(3)}}$ & $\mathrm{SO(7)}^+$  \\
  \null & \null & \null & \null \\
  \hline
   \null & \null & \null & \null \\
  1 & $\mathrm{N^{pqr}}$ & $\frac{\mathrm{SU(3))\times U(1)}}{\mathrm{U(1) \times U(1) }}$ & $\mathrm{SO(7)}^+$  \\
   \null & \null & \null & \null \\
  \hline
  \end{tabular}}
  \caption{The homogeneous $7$-manifolds that admit just one Killing spinors are the squashed $7$-sphere and the infinite
  family of $\mathrm{N^{pqr}}$ manifolds for $pqr \ne 010$. }\label{n=1casi}
\end{table}
\vskip 0.2cm
\eject\vfill
\section{The $\OSp (\mathcal{N}|4)$ supergroup, its superalgebra and its supercosets}
\label{OSPsection}
The key ingredients in the construction of the supergauge completion
of $\mathrm{AdS_4} \times \mathcal{G}/\mathcal{H}$ are provided by
supercoset manifolds of the supergroup $\OSp (\mathcal{N}|4)$ 
\cite{castdauriafre,heidenreich,D'Auria:1982ij,frenico,multanna}. 
For
this reason we dedicate this section to an in depth analysis of such
a supergroup to the structure of its superalgebra described by
appropriate Maurer Cartan equations and to the explicit construction
of coset representatives for relevant instances of supercosets
of the form $\OSp (\mathcal{N}|4)/H$. This lore will be crucial in
our subsequent discussions.
\subsection{The superalgebra}
The real form $\osp(\mathcal{N}|4)$ of the complex  $\osp(\mathcal{N}|4,\mathbb{C})$ Lie superalgebra which is relevant for the study
of $\mathrm{AdS_4} \times \mathcal{G}/\mathcal{H}$
compactifications is that one where the ordinary Lie subalgebra is the following:
\begin{equation}
\sym(4,\mathbb{R}) \, \times \, \so(\mathcal{N})\, \subset \, \osp(\mathcal{N}|4)
\label{realsuba}
\end{equation}
This is quite obvious because of the isomorphism $\sym(4,\mathbb{R}) \, \simeq
\, \so(2,3)$ which identifies $\sym(4,\mathbb{R})$ with the isometry algebra
of anti de Sitter space. The compact algebra $\so(8)$ is instead the
R-symmetry algebra acting on the supersymmetry charges.
\par
The superalgebra $\osp(\mathcal{N}|4)$ can be introduced as follows: consider the two graded $(4+\mathcal{N}) \times (4+\mathcal{N})$ matrices:
\begin{equation}
  \begin{array}{ccccccc}
    \widehat{C} & = & \left( \begin{array}{c|c}
      C\, \gamma_5 & 0 \\
      \hline
      0 & - \frac{\rm i}{4\, e} \, \mathbf{1}_{\mathcal{N}\times \mathcal{N}} \\
    \end{array}\right)  & ; & \widehat{H} & = &\left( \begin{array}{c|c}
      {\rm i} \, \gamma_0\,\gamma_5 & 0 \\
      \hline
      0 & \, - \, \frac{1}{4\, e} \, \mathbf{1}_{\mathcal{N}\times \mathcal{N}} \\
    \end{array} \right) \\
  \end{array}
\label{OmandHmat}
\end{equation}
where $C$ is the charge conjugation matrix in $D=4$. The matrix
$\widehat{C}$ has the property that its upper block is
antisymmetric while its lower one is symmetric. On the other hand, the
matrix $H$ has the property that both its upper and lower blocks are
hermitian. The $\osp(\mathcal{N}|4)$ Lie algebra is then
defined as the set of graded matrices $\Lambda$ satisfying the two
conditions:
\begin{eqnarray}
  \Lambda^T \, \widehat{C} \, + \, \widehat{C} \, \Lambda &
  = & 0 \label{ortosymp}\\
\Lambda^\dagger \, \widehat{H} \, + \, \widehat{H} \, \Lambda &
  = & 0 \label{realsect}
\end{eqnarray}
Eq.(\ref{ortosymp}) defines the complex $\mathrm{osp(\mathcal{N}|4)}$
superalgebra while eq.(\ref{realsect}) restricts it to the
appropriate real section where the ordinary Lie subalgebra is
(\ref{realsuba}).
 The specific form of the matrices $\widehat{C}$ and
$\widehat{H}$ is chosen in such a way that the complete solution of
the constraints (\ref{ortosymp},\ref{realsect}) takes the following
form:
\begin{equation}
  \Lambda \, = \, \left( \begin{array}{c|c}
   -  \ft 14 \, \omega^{ab} \, \gamma_{ab} \, - \,2\, e \, \gamma_a \, \gamma_5 \, E^a \ & \psi_A \\
   \hline
    4 \, {\rm i} \, e \, \overline{\psi}_B \, \gamma_5 & - \, e \,\mathcal{A}_{AB} \
  \end{array} \right)
\label{lambamatra}
\end{equation}
and the Maurer-Cartan equations
\begin{equation}
  d \, \Lambda \, + \, \Lambda \, \wedge \, \Lambda \, = \, 0
\label{Maurocartanosp1}
\end{equation}
read as follows:
\begin{eqnarray}
d \omega^{ab} - \omega^{ac} \, \wedge \,\omega^{db} \, \eta_{cd} + 16 e^2 E^a \,
\wedge \,E^b &=&
-{\rm i}\,  2 e \,  \overline{\psi}_A \, \wedge \gamma^{ab} \gamma^5 \psi_A, \nonumber \\
d E^a - \omega^a_{\phantom{a}c}\, \wedge \, E^c &=& {\rm i} \ft 12 \, \overline{\psi}_A \, \wedge \,\gamma^a
\psi_A, \nonumber\\
d \psi_A - \frac{1}{4} \omega^{ab}\, \wedge \, \gamma_{ab} \psi_A
- e {\mathcal{A}}_{AB} \, \wedge \,\psi_B &=&  2 e \,
 E^a \, \wedge \,\gamma_a \gamma_5 \psi_A, \nonumber\\
d {\mathcal{A}}_{AB} - e  {\mathcal{A}}_{AC}\, \wedge \, {\mathcal{A}}_{CB} &=& 4 \, {\rm i}  \overline{\psi}_A \,
\wedge \,\gamma_5 \psi_B\,.
\label{orfan25}
\end{eqnarray}
Interpreting $E^a$  as the vielbein, $\omega^{ab}$ as the spin connection, and
$\psi^a$ as the gravitino $1$-form, eq.s (\ref{orfan25}) can be
viewed as the structural equations of a supermanifold $\mathrm{AdS}_{4|\mathcal{N}\times 4}$
extending anti de Sitter space with $\mathcal{N}$ Majorana
supersymmetries. Indeed the gravitino $1$--form is a Majorana spinor since, by construction, it satisfies the reality condition
\begin{equation}
  C \, \overline{\psi}_A^T \, = \, \psi_A\,, \quad \quad \overline{\psi}_A \,
  \equiv
\, \psi_A^\dagger \, \gamma_0\,.
\label{majorancondo}
\end{equation}
The supermanifold $\mathrm{AdS}_{4|\mathcal{N}\times 4}$ can be identified with the following supercoset:
\begin{eqnarray}
 \mathcal{M}^{4|4\mathcal{N}}_{osp} & \equiv & \frac{\mathrm{Osp(\mathcal{N} \, | \, 4
})}{\mathrm{SO(\mathcal{N})} \times \mathrm{SO(1,3)}}
\label{ospsuper4_4n}
\end{eqnarray}
Alternatively, the Maurer Cartan equations
can be written in the following more compact form:
\begin{eqnarray}
d \Delta^{xy} + \Delta^{xz} \, \wedge \,\Delta^{ty} \, \epsilon_{zt} &=&
 -\, 4 \, {\rm i}\,  e \,  {\Phi}_A^x \, \wedge \, {\Phi}_A^y, \nonumber \\
d {\mathcal{A}}_{AB} - e  {\mathcal{A}}_{AC}\, \wedge \, {\mathcal{A}}_{CB} &=&
4 \, {\rm i}  {\Phi}_A^x \, \wedge \, {\Phi}_B^y \, \epsilon_{xy}\nonumber\\
d \Phi^x_A \, +  \, \Delta^{xy} \, \wedge \, \epsilon_{yz} \, \Phi^z_A \, - \,
 e \, {\mathcal{A}}_{AB} \, \wedge \,\Phi^x_B &=&  0
\label{orfan26}
\end{eqnarray}
where all $1$-forms are real and, according to the conventions
discussed in appendix \ref{indexconven}, the indices $x,y,z,t$ are
symplectic and take four values. The real symmetric bosonic $1$-form $\Omega^{xy} =
\Omega^{yx}$ encodes the generators of the Lie subalgebra
$\sym(4,\mathbb{R})$, while the antisymmetric real bosonic $1$-form
$\mathcal{A}_{AB}= - \mathcal{A}_{BA}$ encodes the generators of
the Lie subalgebra $\so(\mathcal{N})$. The fermionic
$1$-forms $\Phi^x_A$ are real and, as indicated by their
indices, they transform in the fundamental $4$-dim representation of $\sym(4,\mathbb{R})$
and in the fundamental $\mathcal{N}$-dim representation of
$\so(\mathcal{N})$.
Finally,
\begin{equation}
  \epsilon_{xy}= - \epsilon_{yx} \, = \, \left(\matrix{ 0 & 0 & 0 & 1 \cr 0 & 0 & -1 & 0 \cr 0 & 1 & 0 & 0 \cr
    -1 & 0 & 0 & 0 \cr  } \right)
\label{epsilon}
\end{equation}
is the symplectic invariant metric.
\par
The relation between the formulation (\ref{orfan25}) and
(\ref{orfan26}) of the same Maurer Cartan equations is provided by
the Majorana basis of $d=4$ gamma matrices discussed in appendix
\ref{d4spinorbasis}. Using eq.(\ref{gammareala}), the generators
$\gamma_{ab}$ and $\gamma_a\,\gamma_5$ of the anti de Sitter group
$\mathrm{SO(2,3)}$ turn out to be all given by real symplectic
matrices, as is explicitly shown in eq. (\ref{realgammi}) and the
matrix $\mathcal{C}\,\gamma_5$ turns out to be proportional to
$\epsilon_{xy}$ as shown in eq. (\ref{Chat}). On the other hand a
Majorana spinor in this basis is proportional to a real object times
a phase factor $\exp[ -\, \pi \, {\rm i}\, /\, 4]$.
\par
Hence eq.s (\ref{orfan25}) and eq.s (\ref{orfan26}) are turned ones
into the others upon the identifications:
\begin{equation}
  \begin{array}{ccrcl}
    \Omega^{xy} \, \epsilon_{yz} & \equiv & \Omega^{x}{}_{z} & \leftrightarrow & -  \ft 14 \,
    \omega^{ab} \, \gamma_{ab} \, - \,2\, e \, \gamma_a \, \gamma_5 \, E^a \\
    \null & \null & \mathcal{A}_{AB} & \leftrightarrow & \mathcal{A}_{AB} \\
    \null & \null & \psi_{A}^x & \leftrightarrow & \exp\left[ \ft {- \pi  {\rm i}}{ 4}\right] \, \Phi^x_A \
  \end{array}
\label{conversion}
\end{equation}
As is always the case, the Maurer Cartan equations are just a
property of the (super) Lie algebra and hold true independently of
the (super) manifold on which the $1$-forms are realized: on the supergroup
manifold or on different supercosets of the same supergroup.
\subsection{The relevant supercosets and their relation}
\label{relevsupcos}
We have already introduced the supercoset (\ref{ospsuper4_4n}) which
includes anti de Sitter space and has $4$ bosonic coordinates and
$4\times \mathcal{N}$ fermionic ones. Let us also consider the
following pure fermionic coset:
\begin{eqnarray}
\mathcal{M}^{0|4\mathcal{N}}_{osp} & = & \frac{\mathrm{Osp(\mathcal{N} \, | \, 4
})}{\mathrm{SO(\mathcal{N})} \times \mathrm{Sp(4,\mathbb{R})}}\label{ospsuper0_4n}
\end{eqnarray}
 There is an obvious relation
between these two supercosets that can be formulated in the following
way:
\begin{equation}
   \mathcal{M}^{4|4\mathcal{N}}_{osp} \, \sim \, \mathrm{AdS}_4 \, \times
   \, \mathcal{M}^{0|4\mathcal{N}}_{osp}
\label{relazia}
\end{equation}
In order to explain the actual meaning of eq.(\ref{relazia}) we
proceed as follows. Let the graded matrix $\mathbb{L} \, \in \, \mathrm{Osp(\mathcal{N}|4)}$
be the coset representative of the coset $\mathcal{M}^{4|4 \mathcal{N}}_{osp}$, such that the Maurer Cartan form
$\Lambda$ of eq.(\ref{lambamatra}) can be identified as:
\begin{equation}
  \Lambda = \mathbb{L}^{-1} \, d \mathbb{L}
\label{cosettusrepre}
\end{equation}
Let us now factorize $\mathbb{L}$ as follows:
\begin{equation}
  \mathbb{L} = \mathbb{L}_F \, \mathbb{L}_B
\label{factorL}
\end{equation}
where $\mathbb{L}_F$ is a coset representative for the coset :
\begin{equation}
  \frac{\mathrm{Osp(\mathcal{N} \, | \, 4
})}{\mathrm{SO(\mathcal{N})} \times \mathrm{Sp(4,\mathbb{R})}} \, \ni
\, \mathbb{L}_F
\label{LF}
\end{equation}
and $\mathbb{L}_B$ is the $\mathrm{Osp(\mathcal{N}|4)}$ embedding of a coset
representative of $\mathrm{AdS_4}$, namely:
\begin{equation}
  \mathbb{L}_B \, = \, \left(\begin{array}{c|c}
    \mathrm{L_B} & 0 \\
    \hline
    0 & \mathbf{1}_{\mathcal{N}} \
  \end{array} \right) \quad ; \quad
  \frac{\mathrm{Sp(4,\mathbb{R})}}{\mathrm{SO(1,3)}}\, \ni \,
  \mathrm{L_B}
\label{salamefelino}
\end{equation}
In this way we find:
\begin{equation}
  \Lambda = \mathbb{L}_B^{-1} \, \Lambda_F \, \mathbb{L}_B \, + \, \mathbb{L}_B^{-1}
  \, d \, \mathbb{L}_B
\label{ferrone}
\end{equation}
Let us now write the explicit form of $\Lambda_F $ in analogy to
eq.(\ref{lambamatra}):
\begin{equation}
  \Lambda_F =\left(\begin{array}{c|c}
    \Delta_F & \Theta_A \\
    \hline
    \,  4 \, {\rm i} \, e \, \overline{\Theta}_A \, \gamma_5 & - \, e \, \widetilde{\mathcal{A}}_{AB} \
  \end{array} \right)
\label{fermionform}
\end{equation}
where $\Theta_A$ is a Majorana-spinor valued fermionic $1$-form and
where $\Delta_F$ is an $\sym(4,\mathbb{R})$ Lie algebra valued
$1$-form presented as a $4 \times 4$ matrix. Both $\Theta_A$ as
$\Delta_F$ and $\widetilde{\mathcal{A}}_{AB}$ depend only on the
fermionic $\theta$ coordinates and differentials.
\par
On the other hand we have:
\begin{equation}
  \mathbb{L}_B^{-1}
  \, d \, \mathbb{L}_B \, = \, \left(\begin{array}{c|c}
    \Delta_B & 0 \\
    \hline
    0 & 0 \
  \end{array} \right)
\label{boseforma}
\end{equation}
where the $\Omega_B$ is also an $\sym(4,\mathbb{R})$ Lie algebra valued
$1$-form presented as a $4 \times 4$ matrix, but it depends only on
the bosonic coordinates $x^\mu$ of the anti de Sitter space $\mathrm{AdS_4}$.
Indeed, according to eq(\ref{lambamatra}) we can write:
\begin{equation}
\Delta_B \, = \,    -  \ft 14 \, B^{ab} \, \gamma_{ab} \, - \,2\, e \, \gamma_a \, \gamma_5 \, B^a
\label{Bbwriting}
\end{equation}
where
$\left\{ B^{ab}\, ,\, B^a\right\} $ are respectively the spin-connection and the
vielbein of $\mathrm{AdS_4}$, just as $\left\{ \mathcal{B}^{\alpha\beta}\, ,\, \mathcal{B}^\alpha\right\}
$ are the connection and vielbein of the internal coset manifold
$\mathcal{M}_7$.
\par
Inserting now these results into eq.(\ref{ferrone}) and comparing
with eq.(\ref{lambamatra}) we obtain:
\begin{eqnarray}
\psi_A & = & \mathrm{L_B^{-1} }\, \Theta_A \nonumber\\
\mathcal{A}_{AB} & = & \widetilde{\mathcal{A}}_{AB} \nonumber\\
 -  \ft 14 \, \omega^{ab} \, \gamma_{ab} \, - \,2\, e \, \gamma_a \, \gamma_5 \, E^a
 & = &  -  \ft 14 \, B^{ab} \, \gamma_{ab} \, - \,2\, e \, \gamma_a \, \gamma_5 \,
 B^a \, + \, \mathrm{L_B^{-1} } \, \Delta_F \, \mathrm{L_B }
\label{arcibaldo}
\end{eqnarray}
The above formulae encode an important information. They show how the
supervielbein and the superconnection of the supermanifold
(\ref{ospsuper4_4n}) can be constructed starting from the vielbein
and connection of $\mathrm{AdS_4}$ space plus the Maurer Cartan forms
of the purely fermionic supercoset (\ref{ospsuper0_4n}). In other
words formulae (\ref{arcibaldo}) provide the concrete interpretation
of the direct product (\ref{relazia}). This will also be our starting
point for the actual construction of the supergauge completion in the
case of maximal supersymmetry and for its generalization to the cases
of less supersymmetry.
\subsection{Finite supergroup elements}
We studied the $\osp(\mathcal{N}|4)$
superalgebra but for our purposes we cannot confine ourselves to the
superalgebra, we need also to consider finite elements of the
corresponding supergroup. In particular the supercoset
representative. Elements of the supergroup are described by graded
matrices of the form:
\begin{equation}
  M = \left( \begin{array}{c|c}
    A & \Theta \\
    \hline
   \Pi & D\
  \end{array} \right)
\label{gradamatra}
\end{equation}
where $A,D$ are submatrices made out of even elements of a Grassmann
algebra while $\Theta,\Pi$ are submatrices made out of odd elements
of the same Grassmann algebra. It is important to recall, that the
operations of transposition and hermitian conjugation are defined as
follows on graded matrices:
\begin{eqnarray}
M^T  & = & \left( \begin{array}{c|c}
  A^T & \Pi^T \\
  \hline
 - \,  \Theta^T & D^T
\end{array}\right)  \nonumber\\
M^\dagger  & = & \left( \begin{array}{c|c}
  A^\dagger & \Pi^\dagger \\
  \hline
  \Theta^\dagger & D^\dagger
\end{array} \right)
\label{porfidorosa}
\end{eqnarray}
This is done in order to preserve for the supertrace the same formal
properties enjoyed by the trace of ordinary matrices:
\begin{eqnarray}
\mathrm{Str} \, \left( M\right)  & =  & \mathrm{Tr} \, \left( A\right)  - \mathrm{Tr}\, \left(D\right) \nonumber\\
\mathrm{Str} \, \left( M_1 \, M_2 \right)  & =  & \mathrm{Str} \,
\left(M_2 \, M_1\right)
\label{stracciona}
\end{eqnarray}
Eq.s (\ref{porfidorosa}) and (\ref{stracciona}) have an important
consequence. The consistency of the equation:
\begin{equation}
  M^\dagger = \left( M^T\right) ^\star
\label{starotta}
\end{equation}
implies that the complex conjugate operation on a super matrix must
be defined as follows:
\begin{equation}
  M^\star \, = \, \left( \begin{array}{c|c}
    A^\star & - \Theta^\star \\
    \hline
     \Pi^\star & D^\star \
  \end{array} \right)
\label{staronsupermatra}
\end{equation}
Let us now observe that in the Majorana basis which we have adopted
we have:
\begin{eqnarray}
\widehat{C} & = & {\rm i} \, \, \left(\begin{array}{c|c}
  \epsilon & 0 \\
  \hline
  0 & - \ft 1 {4e} \, \mathbf{1}_{\mathcal{N}\times \mathcal{N}}
\end{array} \right) \, = \, {\rm i} \, \widehat{\epsilon}\nonumber\\
\widehat{H} & =  & \, \left(\begin{array}{c|c}
 {\rm i} \, \epsilon & 0 \\
  \hline
  0 & - \ft 1 {4e} \, \mathbf{1}_{\mathcal{N}\times \mathcal{N}}
\end{array} \right)
\label{porchette}
\end{eqnarray}
where the $4 \times 4$ matrix $\epsilon$ is given by eq.(\ref{Chat}).
Therefore in this basis an orthosymplectic group element $ \mathbb{L} \, \in
\, \OSp(\mathcal{N}|4)$ which satisfies:
\begin{eqnarray}
\mathbb{L}^T \, \widehat{C} \, \mathbb{L}  & = & \widehat{C} \label{reala1}\\
\mathbb{L}^\dagger \, \widehat{H} \, \mathbb{L}  & = & \widehat{H}
\label{reala2}
\end{eqnarray}
has the following structure:
\begin{equation}
  \mathbb{L} \, = \, \left( \begin{array}{c|c}
    \mathcal{S} & \exp \left[-\, {\rm i} \ft {\pi}{4} \right]\Theta \\
    \hline
    \exp \left[-\, {\rm i} \ft {\pi}{4} \right]\,  \Pi & \mathcal{O} \
  \end{array}\right)
\label{struttamatra}
\end{equation}
where the bosonic sub-blocks $\mathcal{S},\mathcal{O}$ are
respectively $4 \times 4 $ and $\mathcal{N}\times \mathcal{N}$ and
real, while the fermionic ones $\Theta,\Pi$ are respectively
$4\times \mathcal{N}$ and $ \mathcal{N}\times 4 $ and also real.
\par
The orthosymplectic conditions (\ref{reala1}) translate into the
following conditions on the sub-blocks:
\begin{eqnarray}
\mathcal{S}^T \, \epsilon \, \mathcal{S} & = &\epsilon - \, {\rm i}\, \ft {1}{4e} \, \Pi^T \, \Pi \nonumber\\
\mathcal{O}^T \,  \mathcal{O} & = & \mathbf{1} \, +  \, {\rm i} \, 4e \,\Theta^T \, \epsilon \,\Theta \nonumber\\
\mathcal{S}^T \, \epsilon \, \Theta & = & -  \, \ft {1}{4e} \, \Pi^T
\, \mathcal {O}
\label{treconde}
\end{eqnarray}
As we see, when the fermionic off-diagonal sub-blocks are zero the
diagonal ones are respectively a symplectic and an orthogonal matrix.
\par
If the graded matrix $\mathbb{L}$ is regarded as the coset
representative of either one of the two supercosets
(\ref{ospsuper4_4n},\ref{ospsuper0_4n}), we can evaluate the explicit
structure of the left-invariant one form $\Lambda$. Using the
$\mathcal{M}^{0|4\times \mathcal{N}}$ style of the Maurer Cartan equations (\ref{orfan26})
we obtain:
\begin{equation}
  \Lambda \, \equiv \, \mathbb{L}^{-1} \, d\mathbb{L} \, = \, \left( \begin{array}{c|c}
    \Delta & \exp \left[- {\rm i } \ft {\pi}{4}\right] \, \Phi \\
    \hline \\
    -   4 e \, \exp \left[ - {\rm i } \ft {\pi}{4}\right] \Phi^T \, \epsilon &  - \, e \, \mathcal{A} \
  \end{array}\right)
\label{ortosymLI}
\end{equation}
where the $1$-forms $\Delta$, $\mathcal{A}$ and $\Phi$ can be
explicitly calculated, using the explicit form of the inverse coset
representative:
\begin{equation}
  \mathbb{L}^{-1} \, = \, \left(\begin{array}{c|c}
    - \epsilon \, \mathcal{S}^T \,  \epsilon & \, \exp \left[- {\rm i } \ft {\pi}{4}\right] \, \ft 1{4e} \,
    \epsilon \Pi^T \, \\
    \hline
   - \exp \left[- {\rm i } \ft {\pi}{4}\right]\, 4e \, \Theta^T \, \epsilon &  \mathcal{O}^T \
  \end{array} \right)
\label{lugubre}
\end{equation}
\begin{eqnarray}
e \mathcal{A} & = & - \, \mathcal{O}^T \, d\mathcal{O} \, - \,{\rm i} \,  4e \, \Theta^T \, \epsilon \, d\Theta \nonumber\\
\Omega  & = & - \, \epsilon \, \mathcal{S}^T \, \epsilon \,
d\mathcal{S}\, - \, {\rm i} \, \ft {1}{4e} \, \Pi^T \, d \Pi \nonumber\\
\Phi & = & \, - \,
\epsilon \, S^T \, \epsilon \, d\Theta \, + \, \ft{1}{4e}
\, \epsilon \, \Pi^T \, d \mathcal{O}
\label{Mcforme}
\end{eqnarray}
\subsection{The coset representative of
$\OSp(\mathcal{N}|4)/\mathrm{Sp(4)} \times \mathrm{SO}(\mathcal{N})$}
It is fairly simple to write an explicit form for the coset
representative of the fermionic supermanifold
\begin{equation}
 \mathcal{M}^{0|4\times \mathcal{N}} \, = \,  \frac{\OSp(\mathcal{N}|4)}{\mathrm{Sp(4,\mathbb{R})} \times \mathrm{SO}(\mathcal{N})}
\label{supermanifoldamente}
\end{equation}
by adopting the upper left block components $\Theta$ of the
supermatrix (\ref{struttamatra}) as coordinates. It suffices to solve
eq.s(\ref{treconde}) for the sub blocks
$\mathcal{S},\mathcal{O},\Pi$. Such an explicit solution is provided
by setting:
\begin{eqnarray}
\mathcal{O}(\Theta) & = & \left(\mathbf{1} \, + \, 4\,{\rm i}\, e \, \Theta^T \, \epsilon \, \Theta\right) ^{1/2} \nonumber\\
\mathcal{S}(\Theta) & = & \left(\mathbf{1} \, + \, 4\,{\rm i}\, e \, \Theta \,  \Theta^T \, \epsilon \,\right) ^{1/2} \nonumber\\
\Pi & = & {4e}  \left(\mathbf{1} \, + \, 4\,{\rm i}\, e \, \Theta^T \, \epsilon \, \Theta\right) ^{-1/2}
\, \Theta^T \, \epsilon \, \left(\mathbf{1} \, + \, 4\,{\rm i}\, e \, \Theta \,  \Theta^T \, \epsilon \,\right)
^{1/2}\nonumber\\
& = & {4e} \, \Theta^T \, \epsilon \null
\label{explicitone}
\end{eqnarray}
In this way we conclude that the coset representative of the
fermionic supermanifold (\ref{supermanifoldamente}) can be chosen to
be the following supermatrix:
\begin{equation}
  \mathbb{L}\left(\Theta \right)  \, = \, \left( \begin{array}{c|c}
    \left(\mathbf{1} \, + \, 4\,{\rm i}\, e \, \Theta \,  \Theta^T \, \epsilon \,\right) ^{1/2} &
    \exp \left[-\, {\rm i} \ft {\pi}{4} \right]\Theta \\
    \hline
   -\, \exp \left[-\, {\rm i} \ft {\pi}{4} \right]\,  {4e} \, \Theta^T \, \epsilon &
    \left(\mathbf{1} \, + \, 4\,{\rm i}\, e \, \Theta^T \, \epsilon \, \Theta\right) ^{1/2}\
  \end{array}\right)
\label{supercosettus}
\end{equation}
By straightforward steps from eq.(\ref{lugubre}) we obtain the inverse of the  supercoset
element (\ref{supercosettus}) in the form:
\begin{equation}
  \mathbb{L}^{-1}\,\left(\Theta \right)  \, = \, \mathbb{L}\,\left(- \, \Theta \right) \, = \, \left( \begin{array}{c|c}
    \left(\mathbf{1} \, + \, 4\,{\rm i}\, e \, \Theta \,  \Theta^T \, \epsilon \,\right) ^{1/2} &
   - \, \exp \left[-\, {\rm i} \ft {\pi}{4} \right]\Theta \\
    \hline
    \exp \left[-\, {\rm i} \ft {\pi}{4} \right]\,  {4e} \, \Theta^T \, \epsilon &
    \left(\mathbf{1} \, + \, 4\,{\rm i}\, e \, \Theta^T \, \epsilon \, \Theta\right) ^{1/2}\
  \end{array}\right)
\label{supercosettusminus}
\end{equation}
Correspondingly we work out the explicit expression  of the Maurer Cartan forms:
\begin{eqnarray}
e \mathcal{A} & = &  \,  \left(\mathbf{1} \, + \, 4\,{\rm i}\, e \, \Theta^{T} \,  \epsilon \,  \Theta \right) ^{1/2} \, d
 \left(\mathbf{1} \, + \, 4\,{\rm i}\, e \,  \Theta^T \, \epsilon \, \Theta \, \right) ^{1/2}
 \, - \,{\rm i} \,  4e \, \Theta^T \, \epsilon \, d\Theta \nonumber\\
\Phi & = &  \left(\mathbf{1} \, + \, 4\,{\rm i}\, e \,  \Theta \,  \Theta^T \,\epsilon \right) ^{1/2} \,  d\Theta \, + \, \Theta \,
d \left(\mathbf{1} \, + \, 4\,{\rm i}\, e \,  \Theta^T \, \epsilon \, \Theta
\right)^{1/2} \nonumber\\
\Delta & = & \left(\mathbf{1} \, + \, 4\,{\rm i}\, e \,  \Theta \,
\Theta^T \epsilon \,
\right)^{1/2} \, d \, \left(\mathbf{1} \, + \, 4\,{\rm i}\, e \,  \Theta \, \Theta^T \epsilon \,
\right)^{1/2} \, - \, {\rm i} \, 4e \, \Theta \, d\Theta^T \, \epsilon
\label{Mcforme2}
\end{eqnarray}


\subsection{Gauged Maurer Cartan $1$-forms of $\OSp(8|4)$ }
A fundamental ingredient in the construction of gauged supergravities
is constituted by the gauging of Maurer Cartan forms of the scalar
coset manifold $\mathrm{G/H}$ (see for instance \cite{parislectures}
for a survey of the subject). The vector fields present in the supermultiplet, which are $1$-forms
defined over the space-time manifold $\mathcal{M}_4$ , are used to deform the Maurer
Cartan $1$-forms of the scalar manifold $\mathrm{G/H}$ that are
instead sections of $T^\star\left( \mathrm{G/H}\right)$. \textit{Mutatis mutandis}, a similar
construction turns out to be quite essential in the problem of gauge completion under consideration.
In our case what will be gauged are the Maurer Cartan $1$-forms of
the supercoset (\ref{ospsuper0_4n}) which contains the fermionic
coordinates of the final superspace we desire to construct. The role
of the space-time gauge fields is instead played by the
$\mathbf{U}$-connection (\ref{so8Ucon1}) of the $\so(8)$ spinor bundle constructed
over the internal $7$-manifold $\left( \mathrm{G/H}\right)_7$.
\par
Accordingly we define:
\begin{equation}
  \widehat{\Lambda} \, \equiv \, \mathbb{L}^{-1} \, \nabla
  \,\mathbb{L} \, = \, \mathbb{L}^{-1} \, \left( d \, \mathbb{L}\, + \,
  \left[ \widehat{\mathbf{U}} \, , \, \mathbb{L}\right]  \right)
\label{gaugedforms}
\end{equation}
where $ \widehat{\mathbf{U}}$ is the supermatrix defined by the
canonical immersion of the $\so(8)$ Lie algebra into the
orthosymplectic superalgebra:
\begin{eqnarray}
\widehat{\mathbf{U}} & = & \left( \begin{array}{c|c}
  0 & 0 \\
  \hline
  0 & \mathbf{U}
\end{array}\right)  \, = \, \mathcal{I} \left( \mathbf{U}\right) \, \nonumber\\
\mathcal{I} & : & \so(8) \, \mapsto \, \osp(8|4)
\label{immersione}
\end{eqnarray}
As a result of their definition, the gauged Maurer Cartan forms
satisfy the following deformed Maurer Cartan equations:
\begin{eqnarray}
  \nabla\widehat{\Lambda} \, + \, \widehat{\Lambda} \, \wedge \, \widehat{\Lambda} &
  =&
   \mathbb{L}^{-1}\left(\Theta \right)  \, \left[ \widehat{F[\mathbf{U}]}
   \, , \,
  \, \mathbb{L}\left(\Theta \right)\right] \label{MCLambdone}
  \end{eqnarray}
  where
  \begin{eqnarray}
  \widehat{F[\mathbf{U}]} & = & \left( \begin{array}{c|c}
  0 & 0 \\
  \hline
  0 & F[\mathbf{U}]
\end{array}\right)
\label{Lambdone2}
\end{eqnarray}
By explicit evaluation, from eq.(\ref{MCLambdone}) we obtain
 the following deformation of the Maurer Cartan equations (\ref{orfan26}):
\begin{eqnarray}
d \widehat{\Delta}^{xy} + \widehat{\Delta}^{xz} \, \wedge \,\widehat{\Delta}^{ty} \, \epsilon_{zt}
 +\, 4 \, {\rm i}\,  e \,  \widehat{{\Phi}}_A^x \, \wedge \, \widehat{{\Phi}}_A^y, & = & \, -\, {\rm i} \, \Theta^x_A \, F_{AB}[\mathbf{U}] \, \Theta_B^y\nonumber \\
\nabla \widehat{{\mathcal{A}}}_{AB} - e  \widehat{{\mathcal{A}}}_{AC}\, \wedge \, \widehat{{\mathcal{A}}}_{CB}
\, - \, 4 \, {\rm i}  \widehat{{\Phi}}_A^x \, \wedge \, \widehat{{\Phi}}_B^y \, \epsilon_{xy} & = & \mathcal{O}_{AP}(\Theta) \, F_{PQ}[\mathbf{U}] \, \mathcal{O}_{QB}(\Theta)
\, - \, F_{AB}[\mathbf{U}]
\nonumber\\
d \widehat{\Phi}^x_A \, +  \, \widehat{\Delta}^{xy} \, \wedge \, \epsilon_{yz} \, \widehat{\Phi}^z_A \, - \,
 e \, \widehat{{\mathcal{A}}}_{AB} \, \wedge \,\widehat{\Phi}^x_B &=&
  \Theta_P^x \, F_{PQ}[\mathbf{U}] \, \mathcal{O}_{QA}(\Theta)
\label{syrotki26}
\end{eqnarray}
The above equations will be our main starting point in the discussion
of the supergauge completion for compactifications with less
preserved supersymmetry.
\subsection{Constrained superspace and the supersolvable
parametrization}
In \cite{Dall'Agata:1998wz} it was demonstrated that, in full analogy
with the solvable parametrization of non compact bosonic coset
manifolds, extensively utilized while dealing with the scalar sector of supergravity models, one can introduce also a
supersolvable parametrization of the supermanifold $\mathcal{M}^{4|4\times
\mathcal{N}}_{osp}$ defined in eq.(\ref{ospsuper4_4n}) (see \cite{GoverH,Dall'Agata:1998wz}). 
This latter
is the supergroup manifold of a solvable super Lie subalgebra $SSolv_{4|\mathcal{N}}\, \subset \,
\mathrm{Osp(\mathcal{N}|4)}$. Similarly to the bosonic case the
solvable parametrization of the supermanifold leads to an enormous
simplification of the Maurer Cartan forms since the coset
representative becomes polynomial in its parameters, yet differently
from the bosonic case the supersolvable algebra $SSolv_{4|\mathcal{N}}$ has smaller
dimension than the dimension of the original coset $\mathcal{M}^{4|4\times
\mathcal{N}}_{osp}$. In other words the supergroup manifold:
\begin{equation}
  \mathcal{SM}^{4|2 \times \mathcal{N}} \, \equiv \, \exp \left[ \, SSolv_{4|\mathcal{N}} \, \right]
\label{Smanif}
\end{equation}
does not contain all the $\Theta$-coordinates but only a subset.
Actually as it is implied by the chosen notation, the solvable supergroup
manifold  $\mathcal{SM}^{4|2 \times \mathcal{N}}$ contains just one-half of the thetas, namely $2 \times \mathcal{N}$.
In \cite{Dall'Agata:1998wz} this was interpreted in terms of
$\kappa$-supersymmetry. Indeed it was advocated that starting from
the general $\kappa$-supersymmetric action of the $M2$-brane, one can
localize it on an $\mathrm{AdS_4} \times \mathbb{S}^7$ background in
a form where all $\kappa$-supersymmetry are already gauged-fixed.
This is the form taken by the general action when the Maurer Cartan
forms of $\mathrm{Osp(\mathcal{N}|4)}$ are written in the
supersolvable parametrization. Alternatively one realizes that the
solvable super Lie algebra $SSolv_{4|\mathcal{N}}$ is nothing else
but the $\mathcal{N}$-extended Poincar\'e superalgebra in three-space
time dimensions, \textit{i.e.} on the membrane world-volume, while
the complete $\mathrm{Osp(\mathcal{N}|4)}$ algebra is simply the
superconformal extension of such an algebra. Hence the supermanifold
(\ref{Smanif}) is just the ordinary Poincar\'e superspace for field
theories on the membrane and the used thetas are the superPoincar\'e
ones while those deleted are the parameters of conformal
supersymmetry which can be non linear realized on the Poincar\'e
ones.
\par
Explicitly the supersolvable parametrization works as follows.
We look for a decomposition of the
$\mathrm{Osp(\mathcal{N}\vert 4)}$ algebra of the following form:
\begin{eqnarray}
\mathrm{Osp(\mathcal{N}\vert 4)}\, &=&\, (\mathrm{SO(1,3)}\otimes \mathrm{SO(\mathcal{N})}\otimes {\cal Q})\oplus
SSolv_{4|\mathcal{N}},
\label{defissolv}
\end{eqnarray}
where ${\cal Q}=\left\{Q_-^A\right\} $ is a subset of the fermionic generators defined by a suitable projection
operator $\mathcal{P}_\pm$
\begin{eqnarray}
Q^A_-\,&=&\, {\cal P}_-\cdot Q^A\nonumber \\
Q^A_+\,&=&\, {\cal P}_+\cdot Q^A ,\\
 {\cal P}_\pm^2\, &=&\,{\cal P}_\pm\,\,;\,\,{\cal P}_+\cdot {\cal
P}_-\,=\,0. \nonumber
\end{eqnarray}
The main idea underlying  the
construction rules of the supersolvable algebra
generating $\mathcal{SM}^{4|2 \times \mathcal{N}}$ as well as the solvable algebra generating
anti de Sitter space is that of {\it grading}. The Cartan
generator contained in the coset of $\mathrm{AdS_4}$ defines a partition of the isometry generators
into eigenspaces corresponding to positive, negative or null
eigenvalues
($\mathbf{g}_{(\pm 1)},\,\mathbf{sg}_{(\pm 1/2)},\,\mathbf{sg}_{(0)}$) and the structure of the
solvable and supersolvable algebras ($Solv$ and $SSolv$) is the
following:
\begin{eqnarray}
\mathbf{g}=\mathrm{SO(2,3)}\, \sim \, \mathrm{Sp(4,\mathbb{R})}\, &\rightarrow & \mathbf{g}_{(-1)}\oplus \mathbf{g}_{(0)}\oplus
\mathbf{g}_{(+1)},\nonumber\\
Solv_{4} \, &=&\, \{{\cal C}\}\oplus \mathbf{g}_{(-1)},\nonumber\\
\label{grading1}
\mathbf{sg}=\mathrm{Osp(\mathcal{N}\vert 4)}&\rightarrow &\mathbf{g}_{(-1)}\oplus \mathbf{sg}_{(0)}\oplus
\mathbf{g}_{(+1)}\oplus \mathbf{sg}_{(-1/2)}
\oplus \mathbf{sg}_{(1/2)},\\
\mathbf{sg}_{(0)}\, &=&\,\mathbf{g}_{(0)}\oplus \mathrm{SO(\mathcal{N})},\nonumber\\
SSolv_{4|\mathcal{N}}\, &=&\, \{{\cal C}\}\oplus \mathbf{g}_{(-1)}\oplus \mathbf{sg}_{(-1/2)}, \nonumber
\end{eqnarray}
where $\mathbf{sg}_{(\pm 1/2)} $ represents the grading induced by the Cartan
generator on the fermionic isometries and the eigenspace $\mathbf{sg}_{(+ 1/2)} $ not
entering the construction of $SSolv$ is the space ${\cal Q}=\{Q^A_+\}$ in eq.
(\ref{defissolv}) and generates the special conformal transformations. Moreover these
generators on the chosen solution of the world volume theory, generate the local
$\kappa$-supersymmetry transformations. As shown in \cite{Dall'Agata:1998wz}
the projection operator which singles out the subspaces $\mathbf{sg}_{(\pm
1/2)}$ is simply given in terms of $4D$-gamma matrices as follows:
\begin{eqnarray}
{\cal P}_\pm \, &=&\, \frac{1}{2}(\unity \pm
\gamma^5\gamma^2),\nonumber\\
\mathbf{sg}_{(\pm 1/2)} \, &=&\, \{Q^A_{\pm}\}\, =\, \{{\cal P}_\pm Q^A\}.
\label{projec}
\end{eqnarray}
It is straightforward to verify that such a projection is compatible
with the Majorana condition and it is immediate to solve such a
constraint in the basis of gamma matrices described in appendix
\ref{d4spinorbasis}. Indeed we find:
\begin{equation}
  Q^A_{\pm} \, = \, \left(\begin{array}{c}
    Q_1^A \\
    Q_2^A \\
    \mp Q_2^A \\
    \pm Q_1^A \
  \end{array} \right)
\label{soluzioneC}
\end{equation}
This implies that the corresponding $\Theta$-coordinates have the same
structure:
\begin{equation}
  \Theta^A \, = \, \Theta_+^A \, \oplus \, \Theta^A_- \quad ; \quad  \Theta^A_{\pm} \, = \, \left(\begin{array}{c}
    \Theta_1^{A\pm} \\
    \Theta_2^{A\pm} \\
    \mp \Theta_2^{A\pm} \\
    \pm \Theta_1^{A\pm} \
  \end{array} \right)
\label{Thetastructure}
\end{equation}
Next it can be immediately verified that the projected $\Theta.s$ satisfy the
following constraints:
\begin{eqnarray}
  \Theta^x_A & = & \Theta^x_{A\pm} \label{choice}\\
  & \Downarrow & \nonumber\\
  \Theta_A^x \, \Theta_B^y \, \epsilon_{xy} & = & 0 \quad \mbox{and} \quad \Theta_A^x \, d\Theta_B^y \,
  \epsilon_{xy}\, = \, \Theta_B^x \, d\Theta_A^y \, \epsilon_{xy}
\label{costrettini}
\end{eqnarray}
As explained in the introduction, in this paper we take a different
point of view. Rather then using the solvable parametrization we take
the complete parametrization of the supercosets (either $\mathcal{M}^{4|4\times
N}$ or $\mathcal{M}^{4|4\times
N}$) but we enforce the constraints (\ref{costrettini}) on the
fermionic coordinates cutting out a sixteen dimensional locus in the
$32$-dimensional one. In this way we preserve all the symmetries and
yet we obtain a formidable simplification of the Maurer Cartan forms
which allows to pursue the gauge completion programme to its very
end.
\subsection{Gauged Maurer Cartan forms in constrained superspace}
Let us now consider the consequences of the constraints
(\ref{costrettini}) on the coset representative
(\ref{supercosettus}), the Maurer Cartan forms (\ref{Mcforme2}) and
their gauged counterparts (\ref{gaugedforms}). On the constrained surface we immediately find:
\begin{eqnarray}
\mathcal{O}(\Theta) & = & 1 \nonumber\\
\mathcal{S}(\Theta) & = & \mathbf{1} \, + \, 2 \, {\rm i}\, e \,
\Theta\, \Theta^T \, \epsilon \nonumber\\
\widehat{\mathcal{A}} & = & \mathcal{A} \, = \, 0 \nonumber\\
\widehat{\Delta} & = & 2 \, {\rm i}\,\left( \nabla \Theta \,
\Theta^T\, - \, \Theta \, \nabla \Theta^T \right)
\label{robepiatte}
\end{eqnarray}
and the gauged Maurer Cartan equations (\ref{syrotki26}) become:
\begin{eqnarray}
d \widehat{\Delta}^{xy} + \widehat{\Delta}^{xz} \, \wedge \,\widehat{\Delta}^{ty} \, \epsilon_{zt}
 +\, 4 \, {\rm i}\,  e \,  \widehat{{\Phi}}_A^x \, \wedge \, \widehat{{\Phi}}_A^y, & = & \, -\, {\rm i} \, \Theta^x_A \, F_{AB}[\mathbf{U}] \, \Theta_B^y\nonumber \\
0 & = & 0
\nonumber\\
d \widehat{\Phi}^x_A \, +  \, \widehat{\Delta}^{xy} \, \wedge \, \epsilon_{yz} \, \widehat{\Phi}^z_A \, &=&
  \Theta_P^x \, F_{PA}[\mathbf{U}]
\label{syrotki28}
\end{eqnarray}
As we are going to show in the sequel, the above equations enable us to write a complete parametrization of
all the FDA superforms adapted to any background $\mathrm{AdS_4} \times
(\mathcal{G}/\mathcal{H})_7$.
\section{Killing spinors of the $\mathrm{AdS_4}$ manifold}
The next main item for the construction of the supergauge completion
is given by the Killing spinors of anti de Sitter space. Indeed,
in analogy with the Killing spinors of the internal $7$-manifold, defined by
eq.(\ref{weakkispinor}) with $m=1$, we can now introduce the notion
of Killing spinors of the $\mathrm{AdS_4}$ space and recognize how they can be
constructed in terms of the coset representative, namely in terms of
the fundamental harmonic of the coset.
\par
The analogue of eq.(\ref{weakkispinor}) is given by:
\begin{equation}
\nabla^{\mathrm{Sp(4)}}\, \chi_x \, \equiv \, \left( d  \, - \, \ft 14 \, B^{ab}
\, \gamma_{ab} \, - \,2\, e \, \gamma_a \, \gamma_5 \,
 B^a \,\right) \, \chi_x \, = \, 0
\label{d4Killing}
\end{equation}
and states that the Killing spinor is a covariantly constant section
of the $\sym (4,\mathbb{R})$ bundle defined over $\mathrm{AdS_4}$.
This bundle is flat since the vanishing of the $\sym (4,\mathbb{R})$
curvature is nothing else but the Maurer Cartan equation of $\sym (4,\mathbb{R})$
and hence corresponds to the structural equations of the $\mathrm{AdS_4}$
manifold. We are therefore guaranteed that there exists a basis of
four linearly independent sections of such a bundle, namely four linearly
independent solutions of eq.(\ref{d4Killing}) which we can
normalize as follows:
\begin{equation}
  \overline{\chi}_x \, \gamma_5 \, \chi_y \, = \, {\rm i} \, \left(   \mathcal{C} \,
  \gamma_5 \right)_{xy}
\label{normakillo4}
\end{equation}
Let $\mathrm{L_B}$ the coset representative mentioned in eq.(\ref{salamefelino}) and
satisfying:
\begin{equation}
- \, \ft 14 \, B^{ab}
\, \gamma_{ab} \, - \,2\, e \, \gamma_a \, \gamma_5 \,
 B^a \, = \, \Delta_B \, = \, \mathrm{L^{-1}_B} \, d \mathrm{L_B}
\label{salamecanino}
\end{equation}
It follows that the inverse matrix $\mathrm{L^{-1}_B}$ satisfies the
equation:
\begin{equation}
  \left( d \, + \, \Delta_B \right) \, \mathrm{L^{-1}_B} \, = \, 0
\label{euchessina}
\end{equation}
Regarding the first index $y$ of the matrix $\left( \mathrm{L^{-1}_B}\right)^y{}_x$ as the
spinor index acted on by the connection $\Delta_B$ and the second
index $x$ as the labeling enumerating the Killing spinors,
eq.(\ref{euchessina}) is identical with eq.(\ref{d4Killing}) and
hence we have explicitly constructed its four independent solutions.
In order to achieve the desired normalization (\ref{normakillo4}) it
suffices to multiply by a phase factor $\exp \left[{-\rm i} \, \ft 14 \pi
\right]$, namely it suffices to set:
\begin{equation}
  \chi^y_{(x)} \, = \, \exp \left[-{\rm i} \, \ft 14 \pi
\right] \, \left( \mathrm{L^{-1}_B}\right)^y{}_x
\label{killini}
\end{equation}
In this way the four Killing spinors fulfill the Majorana condition.
Furthermore since $\mathrm{L^{-1}_B}$ is symplectic it satisfies the
defining relation
\begin{equation}
   \mathrm{L^{-1}_B} \, \mathcal{C} \,
  \gamma_5 \, \mathrm{L_B} \, = \,  \mathcal{C} \,
  \gamma_5 \,
\label{defirelazia}
\end{equation}
which implies (\ref{normakillo4}).
\section{Supergauge completion in mini superspace}
\label{superextension}
As it was observed many years ago in
\cite{fermionMassSpectrum,universal} and it is reviewed at length in
the book \cite{castdauriafre},
given a bosonic Freund Rubin compactification of M-theory on an
internal coset manifold $\mathcal{M}_7 \, = \,\frac{\mathcal{G}}{\mathcal{H}}$ which
admits $\mathcal{N}$ Killing spinors it is fairly easy to extend it consistently to a
\textit{mini-superspace} $ \mathcal{M}^{11|4 \times \mathcal{N}}$
which contains all of the eleven bosonic coordinates but only $4
\times \mathcal{N} $ $\theta$.s, namely those which are associated
with unbroken supersymmetries. We review this extension reformulating
it in such a way that it is suitable
for its generalization to all $\theta$.s namely also
to those associated with broken supersymmetries.
\par
In the original formulation, the mini superspace is viewed as the
following tensor product
\begin{eqnarray}
  \mathcal{M}^{11|4 \times \mathcal{N}} & \equiv & \mathcal{M}^{4|4\mathcal{N}}_{osp} \,
  \times \, \frac{\mathcal{G}}{\mathcal{H}}
\label{minisuper}
\end{eqnarray}
and in order to construct the FDA $p$--forms, in addition to the
Maurer Cartan forms of the above coset, we just need
to introduce the Killing spinors of the bosonic internal manifold.
Let $\eta^A$ be an orthonormal basis of $\mathcal{N}$ eight component Killing
spinors satisfying the Killing spinor condition (\ref{kilspineq7dim}) and the normalization:
\begin{equation}
 \left( \eta^{\underline{A}} \right)^T \, \eta^{\underline{B}} \, = \, \delta^{\underline{AB}}
\label{etaAetaB}
\end{equation}
Next, following \cite{castdauriafre} and \cite{Dall'Agata:1998wz}, whose
results were also summarized in \cite{antonpietro}, we can now write
the complete solution  for the background fields in the case of $\mathrm{AdS_4}
\times \frac{\mathcal{G}}{\mathcal{H}}$ Freund-Rubin backgrounds :
\begin{equation}
  \begin{array}{rcl}
    \widehat{V}^{\underline{a}} & = & \left \{ \begin{array}{rcl}
      \widehat{V}^a & = & E^a \\
      \widehat{V}^\alpha & = & \mathcal{B}^\alpha \, - \, \ft 18  \,
      \overline{\eta}_{\underline{A}} \, \tau^\alpha \, \eta_{\underline{B}} \,
      {\mathcal{A}}_{AB}\\
    \end{array} \right.\\
\widehat{\omega}^{\underline{ab}} & = & \left \{ \begin{array}{rcl}
      \widehat{\omega}^{ab} & = & {\omega}^{ab} \\
      \widehat{\omega}^{\alpha b} & = & 0 \\
      \widehat{\omega}^{\alpha \beta} & = & \mathcal{B}^{\alpha\beta} \,
      + \, \ft e4  \,\overline{\eta}_{\underline{A}} \, \tau^{\alpha\beta} \, \eta_{\underline{B}} \,
      {\mathcal{A}}_{\underline{AB}}\\
    \end{array} \right.\\
    \widehat{\Psi} & = &  \, \eta_{\underline{A}} \, \otimes \, \psi_{\underline{A}} \,
  \end{array}
\label{supervielbein}
\end{equation}
where $\left\{ \mathcal{B}^{\alpha\beta} \, , \,  \mathcal{B}^\alpha\right\}
$ are the spin connection and the vielbein, respectively, of the bosonic seven
dimensional coset manifold $\frac{\mathcal{G}}{\mathcal{H}}$.
\par
Let us now observe that in this formulation of the superextension, the fermionic
coordinates are
actually attached to the space-time manifold $\mathrm{AdS_4}$, which
is superextended to a supercoset manifold:
\begin{equation}
  \mathrm{AdS}_4 \, \stackrel{\mbox{superextension}}{\Longrightarrow}\, \frac{\mathrm{Osp(\mathcal{N} \, | \, 4
})}{\mathrm{SO(\mathcal{N})} \times \mathrm{SO(1,3)}} \, \equiv \,
\mathcal{M}^{4|4\times \mathcal{N}}
\label{superestensione1}
\end{equation}
At the same time the internal manifold $\mathcal{M}_7 \, = \, \frac{\mathcal{G}}{\mathcal{H}}$
is regarded as purely bosonic and it is twisted into the fabric of
the Free Differential Algebra through the notion of the Killing
spinors $\eta_A$, defined as covariantly constant sections of the
$\mathrm{SO(8)}$ spinor bundle over $\mathcal{M}_7$.
\par
Yet whether supersymmetries are preserved or broken precisely depends
on the structure of the $\mathrm{SO(8)}$ spinor bundle on
$\mathcal{M}_7$.
Henceforth it is suggestive to think that the fermionic coordinates
should not be attached to either the internal or to external
manifold, rather they should live as a fiber over the bosonic manifolds.
The first step in order to realize such a programme consists of a
reformulation of the superextension in minisuperspace that treats the
space-time manifold $\mathrm{AdS_4}$ and the internal manifold $\mathcal{M}_7$
in a symmetric way and in both instances relies on the notion of
Killing spinors of the bosonic submanifold as a way of including the
fermionic one. This can be easily done in view of eq.(\ref{relazia})
whose precise meaning we have explained in section \ref{relevsupcos}.
Indeed in view of eq.(\ref{relazia})
we can look at at eq.(\ref{minisuper})
in the following equivalent, but more challenging fashion:
\begin{eqnarray}
  \mathcal{M}^{11|4 \times \mathcal{N}} & = &
{\mathrm{AdS}_4} \, \times \,
  {\mathcal{M}^{0 | 4 \times \mathcal{N}}} \, \times \,
{\mathcal{M}_7}
\nonumber\\
 & \equiv & \underbrace{\frac{\mathrm{Sp(4,\mathbb{R})}}{\mathrm{SO(1,3)}}}_{\mathrm{AdS_4}} \, \times \,
  \underbrace{\frac{\mathrm{Osp(\mathcal{N} \, | \, 4
})}{\mathrm{SO(\mathcal{N})} \times \mathrm{Sp(4,\mathbb{R})}}}_{4 \times \mathcal{N}\,
\mbox{\small fermionic manifold}} \, \times \,
\underbrace{\frac{\mathcal{G}}{\mathcal{H}}}_{\mathcal{M}_7}
\label{minisuperrevi}
\end{eqnarray}
The above equation simply corresponds to the rewriting of eq.(\ref{supervielbein})
in the following way
\begin{equation}
  \begin{array}{rcl}
    \widehat{V}^{\underline{a}} & = & \left \{ \begin{array}{rcl}
      \widehat{V}^a & = & B^a \, - \, \frac{1}{8\, e} \, \overline{\chi}_x \, \gamma^a \, \chi_y \,
      \Delta_F^{xy} \\
      \widehat{V}^\alpha & = & \mathcal{B}^\alpha \, - \, \ft 18  \,
      \overline{\eta}_{\underline{A}} \, \tau^\alpha \, \eta_{\underline{B}} \,
      {\mathcal{A}}_{\underline{AB}}\\
    \end{array} \right.\\
\widehat{\omega}^{\underline{ab}} & = & \left \{ \begin{array}{rcl}
      \widehat{\omega}^{ab} & = & {B}^{ab}\, + \, \ft 12 \, \overline{\chi}_x \, \gamma_5 \, \gamma^{ab} \,
      \chi_y \, \Delta_F^{xy} \\
      \widehat{\omega}^{\alpha b} & = & 0 \\
      \widehat{\omega}^{\alpha \beta} & = & \mathcal{B}^{\alpha\beta} \,
      + \, \ft e4  \,\overline{\eta}_{\underline{A}} \, \tau^{\alpha\beta} \,
      \eta_{\underline{B}} \,
      {\mathcal{A}}_{\underline{AB}}\\
    \end{array} \right.\\
    \widehat{\Psi} & = &   \eta_{\underline{A}} \, \otimes \, \chi_x \, \Phi^{x|\underline{A}}\,
  \end{array}
\label{supervielbein2}
\end{equation}
\section{Gauge completion in the full constrained superspace}
\label{superextensionVERA}
We are now in a position to write an ansatz which solves the
rheonomic parametrization of the FDA curvatures for any $\mathrm{AdS_4} \times \left( \mathcal{G}/\mathcal{H}\right)
_7$ back ground and involves all the $\Theta$--coordinates although
constrained. The extension to mini-superspace provided by eq.s(\ref{supervielbein2})
is our starting point. In those equations the Maurer Cartan forms are
those (\textit{ungauged}) of the supermanifold:
\begin{equation}
  \frac{\mathrm{Osp(\mathcal{N}|4)}}{\mathrm{Sp(4,\mathbb{R})}\times \mathrm{SO(\mathrm{N})}}
\label{Nmanifoldo}
\end{equation}
and therefore are written in terms of $4\times \mathcal{N}$
\textit{unconstrained} fermionic coordinates. The summation on the
indices $\underline{A,B,C}$ is on $\mathcal{N}$--values since
$\eta_{\underline{A}}$ are just the Killing spinors. The MC-forms are
ungauged because, by definition, there is no $\mathbf{U}$-connection in the
directions spanned by the Killing spinors.
\par
The new solution in complete constrained superspace has the following
form:
\begin{equation}
  \begin{array}{rcl}
    \widehat{V}^{\underline{a}} & = & \left \{ \begin{array}{rcl}
      \widehat{V}^a & = & B^a \, - \, \frac{1}{8\, e} \, \overline{\chi}_x \, \gamma^a \, \chi_y \,
      \widehat{\Delta}_F^{xy} \\
      \widehat{V}^\alpha & = & \mathcal{B}^\alpha \, - \, \ft 18  \,
      \overline{\zeta}_{{A}} \, \tau^\alpha \, \zeta_{{B}} \,
      {\widehat{\mathcal{A}}}_{{AB}}\, = \, \mathcal{B}^\alpha \\
    \end{array} \right.\\
\widehat{\omega}^{\underline{ab}} & = & \left \{ \begin{array}{rcl}
      \widehat{\omega}^{ab} & = & {B}^{ab}\, + \, \ft 12 \, \overline{\chi}_x \, \gamma_5 \, \gamma^{ab} \,
      \chi_y \, \widehat{\Delta}_F^{xy} \\
      \widehat{\omega}^{\alpha b} & = & \Delta \omega^{\alpha b} \\
      \widehat{\omega}^{\alpha \beta} & = & \mathcal{B}^{\alpha\beta} \,
      + \, \ft e4  \,\overline{\zeta}_{{A}} \, \tau^{\alpha\beta} \,
      \zeta_{{B}} \,
      {\widehat{\mathcal{A}}}_{{AB}}\, = \, \mathcal{B}^{\alpha\beta} \, + \, \Delta\omega^{\alpha\beta}\\
    \end{array} \right.\\
    \widehat{\Psi} & = &   \zeta_{{A}} \, \otimes \, \chi_x \, \widehat{\Phi}^{x|{A}}\,
  \end{array}
\label{supervielbein3}
\end{equation}
The modifications that have occurred with respect to
eq.(\ref{supervielbein2}) are the following ones:
\begin{enumerate}
  \item The indices $A,B,C$ run on $8$-values and rather then the
  Killing spinors $\eta_{A}$ we have a complete basis of sections $\zeta_A$ of
  the $\so(8)$ spin bundle.
  \item The MC forms  are those of the supermanifold
\begin{equation}
  \frac{\mathrm{Osp({8}|4)}}{\mathrm{Sp(4,\mathbb{R})}\times \mathrm{SO(8)}}
\label{8manifoldo}
\end{equation}
but they are not the ordinary ones, $\mathcal{A},\Delta, \Phi$, rather those gauged by means of
the $\mathbf{U}$-connection on the
$\so(8)$-spinor bundle over $\mathcal{G}/\mathcal{H}$. This is signaled by the hat: $\widehat{\mathcal{A}},\widehat{\Delta},
\widehat{\Phi}$.
  \item The $32$ coordinates of the supermanifold (\ref{8manifoldo})
  are not free, rather they are subject to the constraints
  (\ref{costrettini}). This implies in particular that
  $\widehat{\mathcal{A}}$ vanishes.
  \item The spin connection contains a correction term which is due to the gauging and which we easily calculate below. In particular
  due to this correction the mixed components  $\omega^{\alpha b}$ are no longer
  zero.
\end{enumerate}
It is fairly easy to verify by direct evaluation that the ansatz (\ref{supervielbein3})
verifies the torsion equation (\ref{torsequa}) and the gravitino
equation (\ref{rhoequa}). The mixed part of the spin connection is
just a consequence of the F-deformation of the Maurer Cartan equation
appearing in the first of eq.s(\ref{syrotki28}). By explicit
evaluation we find that without introducing the correction
$\Delta\omega^{\underline{ab}}$ the torsion is not zero, rather it is
given by:
\begin{equation}
  T^{\mathrm{a}} \, = \, \left\{ \begin{array}{rcl}
    T^a & = & \mbox{cost} \, \overline{\chi}_x \, \gamma^a \chi_y \, \zeta_A \, \tau_{\rho\sigma} \zeta_B \, \Theta_A^x \, \Theta^y_B \, \mathcal{C}^{\rho\sigma}{}_{\alpha\beta}
    \, \mathcal{B}^\alpha \, \wedge \, \mathcal{B}^\beta\\
    T^\alpha & = & 0 \
  \end{array}\right.
\label{lucardone}
\end{equation}
In view of the parametrization (\ref{supervielbein3}) this means that
the torsion is of the form:
\begin{eqnarray}
  T^{\underline{a}} & = & H^{\underline{a}|\underline{bc}} \,
  V_{b}\, \wedge
  \, V_{\underline{c}}\nonumber\\
H^{a|\beta\gamma} & = & \mbox{cost} \, \overline{\chi}_x \, \gamma^a \chi_y \, \zeta_A \, \tau_{\rho\sigma} \zeta_B \, \Theta_A^x \, \Theta^y_B \, \mathcal{C}^{\rho\sigma}{}_{\alpha\beta}
    \, \nonumber\\
    \mbox{all other components of} \,
    H^{\underline{a}|\underline{bc}}& = & 0
\label{torsionato}
\end{eqnarray}
which can be reabsorbed by the following redefinition of the spin connection:
\begin{eqnarray}
\omega^{\underline{ab}} & \mapsto & \omega^{\underline{ab}} \, + \, \Delta\omega^{\underline{ab}}\nonumber\\
\Delta\omega^{\underline{ab}} & = & -\, \left(H^{\underline{a}|\underline{bc}} \, - \, H^{\underline{b}|\underline{ac}} \, - \, H^{\underline{c}|\underline{ab}}
\right)\, V^{\underline{c}}
\label{correggo}
\end{eqnarray}
\subsection{The $3$--form}
We have found an explicit expression for the supervielbein $V^{\underline{a}}$ , the
gravitino $1$--form $\Psi$ and ant he spin-connection
$\omega^{\underline{ab}}$. In order to complete  the description of the superextension we need also to
provide an expression for the $3$-form $A^{[3]}$. According to the
general definitions of the FDA curvatures eq.(\ref{FDAcompleta}) and
the rheonomic parametrization (\ref{f4equa}) we find that:
\begin{eqnarray}
d\mathbf{A}^{[3]} & = & \mathbf{F}^{[4]} \, - \, \ft 12 \, \overline{\Psi} \, \wedge \, \Gamma_{\underline{ab}} \, \Psi \, \wedge \,
V^{\underline{a}} \, \wedge \, V^{\underline{b}}  \label{A3astratto}\\
& \Downarrow & \nonumber\\
d\mathbf{A}^{[3]} & = & \, e \, \epsilon_{abcd} \, E^a \, \wedge \, E^b \, \wedge
\, E^c \, \wedge \, E^d  \, + \, \ft 12 \, \overline{\chi}_x \,
\gamma_{ab} \, \chi_y \, \, \Phi^x_A \, \wedge \, \Phi^y \, \wedge
E^a \, \wedge \, E^b \nonumber\\
&& + \, \ft 12 \, \overline{\chi}_x\, \chi_y \, \zeta_A \, \tau_{\alpha \beta } \,
\zeta_B \, \Phi^x_A \, \wedge \, \Phi^y_B \, \wedge \, \mathcal{B}^\alpha  \, \wedge \mathcal{B}^\beta \nonumber\\
&& \, + \, \overline{\chi}_x
\, \gamma_a \, \gamma_5 \chi_y \, \zeta_A \, \tau_\beta  \, \zeta_B
\, \Phi^x _A \, \wedge \, \Phi^y_B \, \wedge \, E^a \, \wedge \, \mathcal{B}^\beta
\label{DA3}
\end{eqnarray}
The expression of $dA^{[3]}$ as a $4$--form is completely explicit in
eq.(\ref{DA3}) and by construction it is integrable in the sense that
$d^2\mathbf{A}^{[3]}=0$. One might desire to solve this equation by
finding a suitable expression for $A^{[3]}$ such that eq.(\ref{DA3})
is satisfied. This is not possible in general terms, namely by using
only the invariant constraints (\ref{costrettini}). In order to find
explicit solutions, one needs to use some explicit coordinate system
and some explicit solution of the constraints. For instance using the
solvable parametrization it was shown in paper
(\cite{Dall'Agata:1998wz}) how to write $A^{[3]}$ in the case of the
seven sphere. This analysis could be pursued also for the other
instances of compactifications with less supersymmetry, but it is not in the spirit we have adopted. Here it is just
the constraints what matters, not their explicit solutions.
\par In the main application we have in mind, namely while localizing
the pure spinor BRST invariant action of the supermembrane M2 on such
backgrounds, we can easily avoid all such problems. We simply
substitute the world volume integral of $A^{[3]}$ with:
\begin{equation}
  \int_{WV_3} \, A^{[3]} \, \mapsto \,  \int_{WV_4} \, dA^{[3]}
\label{fillo}
\end{equation}
where the $4$--dimensional integration volume $WV_4$ is such that its
boundary is the original supermembrane world-volume:
\begin{equation}
  \partial \, WV_4 \, = \, WV_3
\label{urtoffo}
\end{equation}
and we circumvent the problem of solving eq.(\ref{DA3}).
\par
With this observation we have concluded our proof that any $\mathrm{AdS_4}
\times \mathcal{G}/\mathcal{H}$ bosonic solution of M-theory field
equations can be explicit gauge completed to a solution in a
constrained superspace containing all the theta variables both
associated with unbroken as with with broken supersymmetries. Such a
superspace extension is just suited for the pure spinor action of the
M2 brane as derived in \cite{antonpietro}.
\section{Conclusions}
The problem addressed in this paper is the supergauge completion of
M-theory backgrounds of the form $\mathrm{AdS_4} \times \left(
\mathcal{G}/\mathcal{H}\right)$. In short this corresponds to
deriving an explicit parametrization of the $p$-forms of M-theory FDA
in terms of all $32$ fermionic coordinates plus the $11$ bosonic
coordinates of the $7$ manifold associated with the chosen manifold
$\mathrm{AdS_4} \times \left(
\mathcal{G}/\mathcal{H}\right)$. The main motivation of solving such
a problem is that the searched parametrization provides the
necessary information in order to convert the general pure spinor
action of the M2 brane derived in \cite{antonpietro} into an explicit form.

Our solution is based on three ingredients: 1) identification of the obstruction 
which breaks supersymmetry in the non-trivial curvature of an $SO(8)$ connection ${\bf U}$ 
over the spinor bundle of the internal manifold $G/H$; 2) the replacement of $Osp(8|4)$ Maurer-Cartan 
forms with their gauged counterparts by means of the $U$-connection; 3) the implementation of 
a quadratic constraint on the $\theta$ coordinates which in particular admits the solvable parametrization 
of supercoset manifold previously discussed in \cite{Dall'Agata:1998wz}. 

It is rather straightforward that the same ingredients can be used for superstrings in the 
less-supersymmetric backgounds of AdS-type. We leave this subject to a forthcoming publication. 
Nevertheless, in the pure spinor formulation, one needs to BRST transform the constraints (\ref{costrettini}) 
into constraints for the pure spinors. We notice that by solving (\ref{costrettini}) we select a set of independent 
$\theta$'s.  Their BRST variation provides a set of unconstrained commuting spinors on which we can still 
impose the pure spinor constraints. In this way we maintain the balance of degrees of freedom needed to 
cancel the conformal central charge. As a last remark, we point out that the target space supersymmetry is 
realized in a non-linear way and therefore the theory will be manifestly supersymmetric. These 
consideration will be presented more extensively in forthcoming publications. 
 

\section*{Acknowledgments}
We thank L. Castellani, A. Ceresole, and M. Trigiante 
for useful comments. P.F. would also like to thank R. Kallosh and A. Van Proeyen for 
preliminary discussions on the problem of gauge completion held in Stanford several years ago.

\newpage
\appendix
\section{Index conventions}
\label{indexconven}
Due to the complexity of the Lie algebra and super Lie algebra
structures which have to be intertwined together into a single fabric
in order to produce our solution of the FDA equations,  we are forced to introduce a plethora of
different notations for different set of indices and in the present
appendix we summarize our index conventions for the reader's benefit.
\par
We distinguish two sets of index conventions: those relative to the
general theory applying to a generic compactification on $\mathrm{AdS_4}
\times \mathrm{M}_7$ and those relative to the specific example $\mathcal{M}_7
= \mathcal{N^{010}}$
\subsection{Index conventions for the general theory}
We recall that all our indices are flat since we systematically use
differential forms. Furthermore we have tried to incorporate
consistently into our framework
the index conventions adopted in the
series of papers
(\cite{fermionMassSpectrum,universal,castdauriafre,su3su2u1,D'Auria:1983vy,Castellani:1983tc,classificoleo}),
dating back to the eighties and relative to the classification and construction of Freund Rubin
compactifications and readopted in the series of papers
(\cite{Billo:2000zr,Fre':1999xp,Fabbri:1999hw,Fabbri:1999mk,n010massspectrum})
relative to the reinterpretation of such solutions into the context
of the $AdS/CFT$ correspondence.
\begin{enumerate}
\item The underlined lower latin indices from the beginning of the alphabet
$\underline{a,b,c},\dots =0,1, \dots , 10$ run on eleven values and span the vector representation of
the $\so(1,10)$ Lie algebra, namely the tangent Lie algebra of $D=11$
M-theory.
\item The lower latin indices from the beginning of the alphabet ${a,b,c},\dots =0,1, \dots , 3$
(without underlining) run on four values and span the vector
representation of the $\so(1,3)$ Lie algebra, namely the tangent Lie
algebra to the $D=4$ space-time, specifically $AdS_4$.
\item The lower case greek indices from the beginning of the alphabet
$\alpha,\beta,\gamma,\dots = 1,\dots,7$ run on seven values and span
the vector representation of the $\so(7)$ Lie algebra namely the
tangent Lie algebra to the internal seven manifold $\mathcal{M}_7$.
\item The capital latin indices ${A,B,C},\dots =1, \dots , 8$ from the beginning of the
alphabet run on eight values and span the  vector representation of
$\so(8)$. They enumerate the members of an orthonormal basis of
sections $\left\{ \zeta_{{A}}\right\} $ of the spinor bundle on
$\mathcal{M}_7$.
\item Slightly modifying the general conventions of
papers \cite{fermionMassSpectrum,fermionMassSpectrum,universal,su3su2u1,D'Auria:1983vy,classificoleo},
the underlined capital latin indices from the beginning of the alphabet
$\underline{A,B,C}, \dots$ run on $\mathcal{N}$ values and are the vector indices
of the subgroup $\mathrm{SO(\mathcal{N})} \subset
\mathrm{Osp(\mathcal{N}|4)}$. They enumerate the members
of an orthonormal basis of Killing spinors $\eta_{\underline{A}}$.
\item Hence we have in general:
\begin{eqnarray}
\underline{a} & = &\left\{\underbrace{  a}_{\mbox{4 values}}\,\, ,\,\,\underbrace{\alpha}_{\mbox{7 values}} \right\} \nonumber\\
{A} & = & \left\{\underbrace{  \underline{A}}_{\mbox{$\mathcal{N}$ values}}\,\, ,\,\,
\underbrace{\overline{B}}_{\mbox{8 -$\mathcal{N}$ values}} \right\}
\label{cucchini}
\end{eqnarray}
\item The lower case latin indices from the end of the alphabet
$x,y,z,t,\dots$ take four values and are symplectic indices in the fundamental representation of
 $\sym(4,R)$. They enumerate the members $\chi_x$ of an orthonormal basis of
 Killing spinors on the manifold $\mathrm{AdS_4}$.
\end{enumerate}
\section{Spinor identities}
In this section we list some spinor identities which are very useful
in deriving various results discussed in the main text.
\subsection{D=7 gamma matrix basis and spinor identities}
\label{d7spinorbasis}
We begin by writing  the explicit form of the $\tau$ matrices used in
the Kaluza-Klein supergravity literature 
\cite{fermionMassSpectrum} and in particular in the
literature concerning the $\mathrm{N^{010}}$ manifold\footnote{Note that there is a change of
basis with respect to the tau matrices used in paper \cite{antonpietro}}.
\par
The Clifford algebra:
\begin{equation}
  \left\{ \tau_\alpha \, , \, \tau_\beta\right\} \, = \, - \delta_{\alpha \beta
  }
\label{taualgebra}
\end{equation}
is satisfied by the following, real, antisymmetric matrices:
{\scriptsize
\begin{eqnarray*}
  \begin{array}{ccccccc}
    \tau_1 & = & \left( \matrix{ 0 & 0 & 0 & 0 & 0 & 0 & 0 & 1 \cr 0 & 0 & 1 & 0 & 0 & 0 & 0 & 0 \cr 0 &
    -1 & 0 & 0 & 0 & 0 & 0 & 0 \cr 0 & 0 & 0 & 0 & 0 & 0 & 1 & 0 \cr 0 & 0 & 0 & 0 & 0 &
    -1 & 0 & 0 \cr 0 & 0 & 0 & 0 & 1 & 0 & 0 & 0 \cr 0 & 0 & 0 & -1 & 0 & 0 & 0 & 0 \cr
    -1 & 0 & 0 & 0 & 0 & 0 & 0 & 0 \cr  }  \right) & ; & \tau_2 & = & \left( \matrix{ 0 & 0 &
    -1 & 0 & 0 & 0 & 0 & 0 \cr 0 & 0 & 0 & 0 & 0 & 0 & 0 & 1 \cr 1 & 0 & 0 & 0 & 0 & 0 & 0 & 0 \cr 0 & 0 & 0 & 0 &
   0 & 1 & 0 & 0 \cr 0 & 0 & 0 & 0 & 0 & 0 & 1 & 0 \cr 0 & 0 & 0 & -1 & 0 & 0 & 0 & 0 \cr 0 & 0 & 0 & 0 &
    -1 & 0 & 0 & 0 \cr 0 & -1 & 0 & 0 & 0 & 0 & 0 & 0 \cr  }\right) \
\end{array}
\end{eqnarray*}
\begin{eqnarray*}
  \begin{array}{ccccccc}
    \tau_3 & = & \left( \matrix{ 0 & 1 & 0 & 0 & 0 & 0 & 0 & 0 \cr
    -1 & 0 & 0 & 0 & 0 & 0 & 0 & 0 \cr 0 & 0 & 0 & 0 & 0 & 0 & 0 & 1 \cr 0 & 0 & 0 & 0 &
    -1 & 0 & 0 & 0 \cr 0 & 0 & 0 & 1 & 0 & 0 & 0 & 0 \cr 0 & 0 & 0 & 0 & 0 & 0 & 1 & 0 \cr 0 & 0 & 0 & 0 & 0 &
    -1 & 0 & 0 \cr 0 & 0 & -1 & 0 & 0 & 0 & 0 & 0 \cr  }\right)  & ; & \tau_4 & = & \left( \matrix{ 0 & 0 & 0 & 0 & 0 & 0 & -1 & 0 \cr 0 & 0 & 0 & 0 & 0 &
    -1 & 0 & 0 \cr 0 & 0 & 0 & 0 & 1 & 0 & 0 & 0 \cr 0 & 0 & 0 & 0 & 0 & 0 & 0 & 1 \cr 0 & 0 &
    -1 & 0 & 0 & 0 & 0 & 0 \cr 0 & 1 & 0 & 0 & 0 & 0 & 0 & 0 \cr 1 & 0 & 0 & 0 & 0 & 0 & 0 & 0 \cr 0 & 0 & 0 &
    -1 & 0 & 0 & 0 & 0 \cr  }\right)  \
\end{array}
\end{eqnarray*}
\begin{eqnarray*}
  \begin{array}{ccccccc}
   \tau_5 & = & \left(\matrix{ 0 & 0 & 0 & 0 & 0 & 1 & 0 & 0 \cr 0 & 0 & 0 & 0 & 0 & 0 & -1 & 0 \cr 0 & 0 & 0 &
    -1 & 0 & 0 & 0 & 0 \cr 0 & 0 & 1 & 0 & 0 & 0 & 0 & 0 \cr 0 & 0 & 0 & 0 & 0 & 0 & 0 & 1 \cr
    -1 & 0 & 0 & 0 & 0 & 0 & 0 & 0 \cr 0 & 1 & 0 & 0 & 0 & 0 & 0 & 0 \cr 0 & 0 & 0 & 0 & -1 & 0 & 0 & 0 \cr  } \right) & ; & \tau_6 & = & \left(
    \matrix{ 0 & 0 & 0 & 0 & -1 & 0 & 0 & 0 \cr 0 & 0 & 0 & 1 & 0 & 0 & 0 & 0 \cr 0 & 0 & 0 & 0 & 0 & 0 &
    -1 & 0 \cr 0 &
    -1 & 0 & 0 & 0 & 0 & 0 & 0 \cr 1 & 0 & 0 & 0 & 0 & 0 & 0 & 0 \cr 0 & 0 & 0 & 0 & 0 & 0 & 0 & 1 \cr 0 & 0 & 1 &
   0 & 0 & 0 & 0 & 0 \cr 0 & 0 & 0 & 0 & 0 & -1 & 0 & 0 \cr  } \right)  \
\end{array}
\end{eqnarray*}
\begin{equation}
  \begin{array}{ccc}
    \tau_7 & = & \left( \matrix{ 0 & 0 & 0 & 1 & 0 & 0 & 0 & 0 \cr 0 & 0 & 0 & 0 & 1 & 0 & 0 & 0 \cr 0 & 0 & 0 & 0 & 0 & 1 & 0 & 0 \cr
    -1 & 0 & 0 & 0 & 0 & 0 & 0 & 0 \cr 0 & -1 & 0 & 0 & 0 & 0 & 0 & 0 \cr 0 & 0 &
    -1 & 0 & 0 & 0 & 0 & 0 \cr 0 & 0 & 0 & 0 & 0 & 0 & 0 & 1 \cr 0 & 0 & 0 & 0 & 0 & 0 & -1 & 0 \cr  } \right)  \
  \end{array}
\label{tauexplicit}
\end{equation}
}
Let $\zeta_{\underline{A}}$ be an orthonormal basis of section for
the spinor bundle on $\mathcal{M}_7$, namely:
\begin{equation}
  \overline{\zeta}_{\underline{A}} \,
   \zeta_{\underline{B}}
  \, = \, \delta_{\underline{AB}}
\label{ortosezioni}
\end{equation}
Now let $Q_{\underline{AB}} \, = \, - Q_{\underline{BA}}$ be any
$\mathrm{SO(8)}$ Lie algebra valued $1$-form and let us define the
following objects:
\begin{eqnarray}
\Delta^{\alpha \beta } & \equiv  & \zeta_{\underline{A}} \tau^{\alpha\beta} \,  \zeta_{\underline{B}} \, Q^{AB}\nonumber\\
\Theta^{\alpha} & \equiv  & \zeta_{\underline{A}} \tau^{\alpha} \,  \zeta_{\underline{B}} \, Q^{AB}\nonumber\\
\Xi^{\alpha} & \equiv  & \zeta_{\underline{A}} \tau^{\alpha} \,  \zeta_{\underline{B}} \, Q^{AC} \, \wedge \, Q^{CD}
\nonumber\\
\Pi^{\alpha\beta} & \equiv  & \zeta_{\underline{A}} \tau^{\alpha\beta}
\,  \zeta_{\underline{B}} \, Q^{AC} \, \wedge \, Q^{CD}
\label{oggettistica}
\end{eqnarray}
Then using the negative metric to saturate the $\so(7)$ vector
indices, as it is appropriate in our conventions, we find the following
identities:
\begin{eqnarray}
\left( -\ft {1}{16} \, \Delta^{\alpha \beta } \, \tau_{\alpha \beta } \, + \, \ft 18 \, \Theta^\alpha  \,
\tau_\alpha\right) \, \zeta_{\underline{A}}  & = & Q_{\underline{AB}} \, \zeta_{\underline{B}} \nonumber\\
\Delta^{\alpha \beta }\, \wedge \, \Theta^{\beta } & = & 4 \,
\Xi^{\alpha} \nonumber\\
- \, \Delta^{\alpha \beta }\, \wedge \, \Delta^{\beta \gamma} & = & - 4 \,
\Pi^{\alpha\beta} \, + \, \Theta^\alpha  \, \wedge \, \Theta^\beta
\label{idiotiche}
\end{eqnarray}
Next we consider the spinor identities in $4$-dimensions.
\subsection{D=4 $\gamma$-matrix basis and spinor identities}
\label{d4spinorbasis}
In this section we construct a basis of $\so(1,3)$ gamma matrices
such that it explicitly realizes the isomorphism $\so(2,3)  \sim
\sym (4,\mathbb{R})$ with the conventions used in the main text.
Naming $\sigma_i$ the standard Pauli matrices:
\begin{equation}
  \sigma_1 =\left( \begin{array}{cc}
    0 & 1 \\
    1 & 0 \
  \end{array}\right) \quad ; \quad \sigma_2 =\left( \begin{array}{cc}
    0 & -{\rm i} \\
    {\rm i} & 0 \
  \end{array}\right) \quad ; \quad \sigma_3 =\left( \begin{array}{cc}
    1 & 0 \\
    0 & -1 \
  \end{array}\right)
\label{paulini}
\end{equation}
we realize the $\so(1,3)$ Clifford algebra:
\begin{equation}
  \left\{ \gamma_a \, , \, \gamma_b \right\} \, = \, 2 \, \eta_{ab}
  \quad ; \quad \eta_{ab} \, = \, \mbox{diag} \left( + , - , - , - \right)
\label{d4clif}
\end{equation}
by setting:
\begin{equation}
  \begin{array}{ccccccc}
    \gamma_0 & = &  \sigma_2 \, \otimes \, \mathbf{1}  & ; & \gamma_1 & = & {\rm i} \, \sigma_3  \,
    \otimes \, \sigma_1\\
    \gamma_2 & = & {\rm i} \sigma_1 \, \otimes \, \mathbf{1} & ; &
    \gamma_3 & = & {\rm i} \sigma_3 \, \otimes \, \sigma_3 \\
    \gamma_5 & = & \sigma_3 \, \otimes \, \sigma_2 & ; & \mathcal{C} & = & {\rm i} \sigma_2 \,
    \otimes \, \mathbf{1}
  \end{array}
\label{gammareala}
\end{equation}
where $\gamma_5$ is the chirality matrix and $\mathcal{C}$ is the
charge conjugation matrix. Making now reference to eq.s
(\ref{OmandHmat}) and (\ref{ortosymp}) of the main text we see that
the antisymmetric matrix entering the definition of the
orthosymplectic algebra, namely $\mathcal{C}\, \gamma_5$ is the
following one:
\begin{equation}
  \mathcal{C}\, \gamma_5 \, = \, {\rm i} \left(\matrix{ 0 & 0 & 0 & 1 \cr 0 & 0 & -1 & 0 \cr 0 & 1 & 0 & 0 \cr
    -1 & 0 & 0 & 0 \cr  } \right)
\label{Chat}
\end{equation}
namely it is proportional, through an overall ${\rm i}$-factor, to a real
completely off-diagonal matrix. On the other hand all the generators
of the $\so(2,3)$ Lie algebra, \textit{i.e.} $\gamma_{ab}$ and
$\gamma_a\, \gamma_5$ are real, symplectic $4 \times 4$ matrices.
Indeed we have
\begin{equation}
  \begin{array}{ccccccc}
    \gamma_{01} & = &\left(  \matrix{ 0 & 0 & 0 & -1 \cr 0 & 0 & -1 & 0 \cr 0 &
    -1 & 0 & 0 \cr -1 & 0 & 0 & 0 \cr  }\right)  & ; & \gamma_{02} & = & \left(
     \matrix{ 1 & 0 & 0 & 0 \cr 0 & 1 & 0 & 0 \cr 0 & 0 &
    -1 & 0 \cr 0 & 0 & 0 & -1 \cr  }\right) \\
    \null & \null & \null & \null & \null & \null & \null \\
    \gamma_{12} & = &\left(  \matrix{ 0 & 0 & -1 & 0 \cr 0 & 0 & 0 & 1 \cr
    -1 & 0 & 0 & 0 \cr 0 & 1 & 0 & 0 \cr  }\right) & ; & \gamma_{13} & = &\left(  \matrix{ 0 & 0 & 0 & -1 \cr 0 & 0 &
    -1 & 0 \cr 0 & 1 & 0 & 0 \cr 1 & 0 & 0 & 0 \cr  }\right)\\
    \null & \null & \null & \null & \null & \null & \null \\
    \gamma_{23} & = & \left(  \matrix{ 0 & 1 & 0 & 0 \cr
    -1 & 0 & 0 & 0 \cr 0 & 0 & 0 & 1 \cr 0 & 0 & -1 & 0 \cr  }\right) & ; & \gamma_{34} & = &\left(
     \matrix{ 0 & 0 & 1 & 0 \cr 0 & 0 & 0 & -1 \cr
    -1 & 0 & 0 & 0 \cr 0 & 1 & 0 & 0 \cr  }\right) \\
    \null & \null & \null & \null & \null & \null & \null \\
    \gamma_{0} \,\gamma_5 & = & \left(  \matrix{ 0 & 0 & 0 & 1 \cr 0 & 0 & -1 & 0 \cr 0 & 1 & 0 & 0 \cr
    -1 & 0 & 0 & 0 \cr  }\right) & ; & \gamma_{1} \,\gamma_5 & = &\left(
    \matrix{ -1 & 0 & 0 & 0 \cr 0 & 1 & 0 & 0 \cr 0 & 0 &
    -1 & 0 \cr 0 & 0 & 0 & 1 \cr  }\right) \\
    \null & \null & \null & \null & \null & \null & \null \\
    \gamma_{2} \,\gamma_5 & = & \left(  \matrix{ 0 & 0 & 0 & -1 \cr 0 & 0 & 1 & 0 \cr 0 & 1 & 0 & 0 \cr
    -1 & 0 & 0 & 0 \cr  }\right)& ; & \gamma_{3} \,\gamma_5 & = &\left(  \matrix{
   0 & 1 & 0 & 0 \cr 1 & 0 & 0 & 0 \cr 0 & 0 & 0 & 1 \cr 0 & 0 &
   1 & 0 \cr  }\right) \
  \end{array}
\label{realgammi}
\end{equation}
On the other hand we find that $\mathcal{C}\gamma_0 = {\rm i} \,
\mathbf{1}$. Hence the Majorana condition becomes:
\begin{equation}
  {\rm i} \, \psi \, = \, \psi^\star
\label{Majorana}
\end{equation}
so that a Majorana spinor is just a real spinor multiplied by an
overall phase $\exp \left [- i \frac \pi 4\right ]$.
\par
These conventions being fixed let $\chi_x$ ($x=1,\dots,4$) be a set of
(commuting) Majorana spinors normalized in the following way:
\begin{equation}
  \begin{array}{lclcl}
    \chi_x & = & \mathcal{C}\, \overline{\chi}_x^T & ; & \mbox{Majorana condition} \\
    \overline{\chi}_x \, \gamma_5 \, \chi_y & = & {\rm i} \, \left( \mathcal{C}\, \gamma_5\right) _{xy} & ; &
    \mbox{symplectic normal basis} \
  \end{array}
\label{festoso}
\end{equation}
Then by explicit evaluation we can verify the following Fierz
identity:
\begin{equation}
  \ft 12 \, \gamma^{ab} \, \chi_z \, \overline{\chi}_x \, \gamma_5 \,
  \gamma_{ab} \, \chi_y \, - \, \gamma_a \, \gamma_5 \, \chi_z \,
  \overline{\chi}_x \, \gamma_a \, \chi_y \, = \, - \, 2{\rm i} \, \left[
  \left( C\gamma_5\right) _{zx} \, \chi_y \, + \, \left( C\gamma_5\right) _{zy} \,
  \chi_x \right]
\label{firzusd4x1}
\end{equation}
Another identity which we can prove by direct evaluation is the
following one:
\begin{eqnarray}
& \overline{\chi}_x \, \gamma_5 \gamma_{ab} \, \chi_y \, \overline{\chi}_z \, \gamma^b \, \chi_t
\, - \, \overline{\chi}_z \, \gamma_5 \gamma_{ab} \, \chi_t \, \overline{\chi}_x \, \gamma^b \, \chi_y
= &\nonumber\\
&  {\rm i}\left( \overline{\chi}_x \, \gamma_a \, \chi_t \, \left(  \mathcal{C}\,
\gamma_5\right)_{yz} \, + \, \overline{\chi}_y \, \gamma_a \, \chi_t \, \left( \mathcal{C}\,
\gamma_5\right)_{xz}  + \overline{\chi}_x \, \gamma_a \, \chi_z \, \left( \mathcal{C}\,
\gamma_5\right)_{yt} \, + \, \overline{\chi}_y \, \gamma_a \, \chi_z \, \left( \mathcal{C}\,
\gamma_5\right)_{xt}\right) &\nonumber\\
\label{firzusd4x2}
\end{eqnarray}
Both these identities are of high relevance in our discussion of the supergauge
completion.
\section{The explicit form of the $\mathbf{U}$--connection in a pair of
examples}
Since the central item in deriving the gauge superextension is
provided by the $\mathbf{U}$--connection on the $\so(8)$ spinor bundle, it is
appropriate to spell out the explicit form of a such a $1$--form at
least in a couple of cases. To this effect we shall consider the
spaces $\mathrm{Q}^{111}$ and $\mathrm{N^{010}}$.
\subsection{The $Q^{111}$ sasakian manifold}
\label{q111}
The $7$ manifold $Q^{111}$ is an $\mathbb{S}^1$ fibration over the
product of three $\mathbb{P}^1$:
\begin{equation}
  Q^{111} \, \stackrel{\pi}{\Longrightarrow} \, \mathbb{P}^1 \,
  \times \, \mathbb{P}^1 \, \times \, \mathbb{P}^1
\label{3pi1}
\end{equation}
the fibration being:
\begin{equation}
   Q^{111} \, \sim \, \mathcal{\mathcal{O}}\left( \mathbb{P}^1,  1\right) \,
   \ \otimes\, \mathcal{\mathcal{O}}\left( \mathbb{P}^1,  1\right) \, \otimes \, \mathcal{\mathcal{O}}\left( \mathbb{P}^1, 1\right)
\label{pqr=111}
\end{equation}
This means that, as a coset manifold, it can be described as the particular instance $(p,q,r) =(1,1,1)$  in the infinite family of homogeneous spaces:
\begin{eqnarray}
  Q^{pqr} & = & \frac{\mathrm{SU(2)_1 \times SU(2)_2 \times SU(2)_3 \times
  U(1)}}{\mathrm{U(1) \times U(1) \times U(1)}}\label{su(2)cubeU(1)}\\
  Z & = & p \, J^3_{(1)} \, + \, q \, J^3_{(2)}\, + \,  r\, J^3_{(2)}
  \, + \, Y \label{Zgenera}
\end{eqnarray}
by definition $Z$ being the Cartan generator that is not in the
subalgebra $\mathrm{H}=\mathrm{U(1)}\times\mathrm{U(1)}\times \mathrm{U(1)}$,
$J^a_{(i)}$ ($a=1,2,3$) being the generators of $\mathrm{SU(2)}_i$
and the hypercharge $Y$ being the generator of $\mathrm{U(1)}$ in the
numerator group $\mathrm{G}$.
\par
These $7$-manifolds were originally introduced in \cite{Q111noi} and
their role as solutions of $D=11$ supergravity was there discussed. In
particular their holonomy and Killing spinors were calculated
explicitly in \cite{Q111noi}, showing that for $(p,q,r) =(1,1,1)$ there is
$\so(8)$-holonomy equal to $\su(3)$ and two Killing spinors, while in
all the other cases all supersymmetries are broken. In the context of
the AdS/CFT correspondence, the algebraic structure of the sasakian
manifolds was shown to determine the form of the dual gauge theories in
\cite{Fabbri:1999hw} and in that paper the gauge dual of
$Q^{111}$ was also derived. Finally the complete Kaluza-Klein
spectrum of M-theory on $AdS_4 \times Q^{111}$ and its organization
in $\mathrm{Osp(2|4) \times SU(2)^3}$ multiplets was derived in
\cite{q111Merlatti}.
We review here the essential steps in the geometrical construction of $Q^{111}$ in
order to calculate the explicit form of the $\so(8)$ connection
\par
We begin by writing the Maurer Cartan equations of the Lie algebra
$\mathbb{G}=\su(2) \oplus \su(2) \oplus \su(2) \oplus \uu(1)$, by
enumerating its generators from one to ten, the first triplet $e_1,e_2,e_3$ being the
generators of the first $\su(2)$, the second triplet $e_4,e_5,e_6$ the generators
of the second $\su(2)$ and so on. The last generator $e_{10}$ is
associated with the abelian $\uu(1)$ algebra. Correspondingly we
have:
\begin{eqnarray}
&&\left. \begin{array}{rcl}
0 & = & de_{1+3 i} \, + \, e_{2+3 i} \, \wedge \, e_{3+3 i} \\
0  &= &  de_{2+3 i} \, - \, e_{1+3 i} \, \wedge \, e_{3+3 i} \\
0  &=&  de_{3+3i} \, + \, e_{1+3 i} \, \wedge \, e_{2+3 i}  \
\end{array} \right \}\quad
i=0,1,2 \nonumber\\
&& 0  = de_{10}
\label{oldmaure}
\end{eqnarray}
Next we perform a change of basis in the above $10$-dimensional
algebra introducing the following new set of $1$-forms:
\begin{eqnarray}
  &&\begin{array}{rclcrcl}
    \Sigma_1 & = & \frac {1}{4\sqrt{2}\, e} e_1 &;&
    \Sigma_2 & = & \frac {1}{4\sqrt{2}\, e} e_2 \\
    \Sigma_3 & = & \frac {1}{4\sqrt{2}\, e} e_4 & ; &
    \Sigma_4 & = & \frac {1}{4\sqrt{2}\, e} e_5 \\
    \Sigma_5 & = & \frac {1}{4\sqrt{2}\, e} e_7 &; &
    \Sigma_6 & = & \frac {1}{4\sqrt{2}\, e} e_8 \
    \end{array}\nonumber\\
&&\begin{array}{rcl}
    \Sigma_7 & = & \frac {1}{8\, e} \left( e_3 \, + \, e_6 \, + \, e_9 \, + \, e_{10} \right)  \\
    \Sigma_8 & = & \frac{1}{2} \left(  e_3 \, - \, e_6 \, - \, e_9 \, + \, e_{10} \right)   \\
    \Sigma_9 & = & \frac{1}{2} \left( - \, e_3 \, + \, e_6 \, - \, e_9 \, + \, e_{10} \right) \\
    \Sigma_{10} & = & \frac{1}{2} \left(- \,  e_3 \, - \, e_6 \, + \, e_9 \, + \, e_{10} \right) \
  \end{array}
\label{cambiobase}
\end{eqnarray}
The meaning of the above rearrangement is the following. Apart from
the rescaling by the factor $\ft{1}{4\sqrt{2}\, e}$ the first six
generators are, two by two, the vielbeins of the three copies of the
$2$-dimensional projective space $\mathbb{P}^1 \sim
\mathrm{SU(2)/U(1)}$. The last four generators correspond to an
orthogonal basis in the space spanned by the four Cartan generators,
such that the first element in the basis is dual to the generator $Z$
of eq.(\ref{Zgenera}) with $p=q=r=1$. In this way $\Sigma_7$ can be
identified as the $7$th-vielbein of $Q^{111}$. The remaining three
$1$-forms $\Sigma_{8,9,10}$ provide a basis for the $\mathrm{H}$-subalgebra
$\mathbb{H}=\uu(1)\oplus\uu(1)\oplus\uu(1)$. The rescalings of the
vielbeins have being chosen in such a way as to produce a diagonal
Ricci tensor with $7$-eigenvalues all equal to $12 e^2$ as it is
required in order for the manifold to be a solution of $D=11$
supergravity. Here as above $e$ denotes the Freund Rubin parameter.
\par
Writing the Maurer Cartan equations (\ref{oldmaure}) in the new basis the Maurer Cartan equations (\ref{oldmaure}) we can  use them  to  calculate the spin connection
$B^{\alpha \beta }$ of the $7$-manifold by setting:
\begin{equation}
  \mathcal{B}^\alpha = \left\{ \Sigma_1 \, , \, \dots \Sigma_7\right\}
\label{vielbine}
\end{equation}
and implementing  the vanishing of the torsion:
\begin{equation}
  d\mathcal{B}^\alpha \, + \, \mathcal{B}^{\alpha \beta } \, \wedge \, \mathcal{B}^\beta \, = \, 0
\label{lucamoria}
\end{equation}
This leads to the calculation of the Riemann tensor and of the Ricci
tensor:
\begin{equation}
  \mathcal{R}^\alpha_\beta = 12 e^2 \, \delta^\alpha_\beta
\label{riccius}
\end{equation}
as required.
\par
The connection on the $\so(8)$-bundle can now be easily calculated.
From its definition:
\begin{equation}
  \mathbf{U} \, = \, - \, \ft 14 \, \mathcal{B}^{\alpha \beta } \,
  \tau_{\alpha \beta } \, - \, e \, \mathcal{B}^\alpha \, \tau_\alpha
\label{Uq111}
\end{equation}
we can obtain its explicit form, provided we use an explicit representation of the
$\tau$-matrices, satisfying the Clifford algebra (\ref{taualgebra}). In appendix
\ref{d7spinorbasis} we displayed an explicit realization of the $\tau$ matrices which is
well adapted to the discussion of the $\mathrm{N^{010}}$ manifold and is particularly simple.
Certainly we could use such a basis also for the $Q^{111}$-manifold, yet, in this case it is
convenient to use another basis $\tau_\alpha^\prime$ , related to the $\tau_\alpha$  of eq.s(\ref{tauexplicit})  by an orthogonal $\mathrm{SO(8)}$ transformation:
\begin{equation}
  \tau_\alpha^\prime \, = \, \mathbf{O}\, \tau_\alpha \, \mathbf{O}^T
\label{otranspa}
\end{equation}
where:
\begin{equation}
 \mathbf{O}\, = \, \left( \matrix{ 0 & -  \frac{1}{2}    & 0 & \frac{1}{2} & 0 & -  \frac{1}{2}    & 0 & \frac{1}
   {2} \cr \frac{1}{2} & 0 & -  \frac{1}{2}    & 0 & \frac{1}{2} & 0 & \frac{1}{2} & 0 \cr \frac{1}
   {2} & 0 & \frac{1}{2} & 0 & \frac{1}{2} & 0 & -  \frac{1}{2}    & 0 \cr 0 & \frac{1}{2} & 0 & \frac{1}
   {2} & 0 & \frac{1}{2} & 0 & \frac{1}{2} \cr \frac{1}{2} & 0 & \frac{1}{2} & 0 & -  \frac{1}{2}
      & 0 & \frac{1}{2} & 0 \cr \frac{1}{2\,{\sqrt{2}}} & \frac{1}{2\,{\sqrt{2}}} & \frac{-1}{2\,{\sqrt{2}}} &
    \frac{1}{2\,{\sqrt{2}}} & \frac{-1}{2\,{\sqrt{2}}} & \frac{-1}{2\,{\sqrt{2}}} & \frac{-1}{2\,{\sqrt{2}}} &
    \frac{-1}{2\,{\sqrt{2}}} \cr 0 & \frac{1}{2} & 0 & -  \frac{1}{2}    & 0 & -  \frac{1}{2}
      & 0 & \frac{1}{2} \cr \frac{1}{2\,{\sqrt{2}}} & \frac{-1}{2\,{\sqrt{2}}} & \frac{-1}{2\,{\sqrt{2}}} & \frac{-1}
   {2\,{\sqrt{2}}} & \frac{-1}{2\,{\sqrt{2}}} & \frac{1}{2\,{\sqrt{2}}} & \frac{-1}{2\,{\sqrt{2}}} & \frac{1}
   {2\,{\sqrt{2}}} \cr  }\right)
\label{Omatra}
\end{equation}
If the $\tau_\alpha$ used in eq.(\ref{Uq111}) are the
$\tau^\prime_\alpha$, defined in eq.(\ref{Omatra}), we get a
block-diagonal structure for the $\mathbf{U}$-matrix:
\begin{equation}
  \mathbf{U} \, = \, \left( \begin{array}{c|c}
    \mathbf{U}_2 & 0 \\
    \hline
    0 & \mathbf{U}_6 \
  \end{array}\right)
\label{bennet}
\end{equation}
where:
\begin{eqnarray}
\so(2) \, \ni \, \mathbf{U}_2 & = & \left(\matrix{ 0 & e\,{{\Sigma }_7} +
   \frac{{{\Sigma }_8}}{4} +
   \frac{{{\Sigma }_9}}{4} +
   \frac{{{\Sigma }_{10}}}{4} \cr - e\,
      {{\Sigma }_7}   -
   \frac{{{\Sigma }_8}}{4} -
   \frac{{{\Sigma }_9}}{4} -
   \frac{{{\Sigma }_{10}}}{4} & 0 \cr  } \right)\nonumber\\
   \null & \null &\null \nonumber\\
   & = & \ft 12 \, \left( \begin{array}{cc}
     0 & e_{10} \\
     -e_{10} & 0 \
   \end{array}\right)
\label{so2conne}
\end{eqnarray}
and
\begin{eqnarray}
\so(6) \, \ni \, \mathbf{U}_6 & =&
\left(\matrix{ 0 & \frac{-{e_4}}{2\,{\sqrt{2}}} &
    \frac{{e_5}}{2\,{\sqrt{2}}} & \frac{{e_7} -
     {e_8}}{4} & \frac{-{e_3} + {e_6} + {e_9}}
   {2} & \frac{-{e_7} - {e_8}}{4} \cr \frac{{e_4}}
   {2\,{\sqrt{2}}} & 0 & \frac{{e_3} + {e_6} -
     {e_9}}{2} & \frac{{e_1} + {e_2}}{4} &
    \frac{-{e_5}}{2\,{\sqrt{2}}} & \frac{{e_1} -
     {e_2}}{4} \cr \frac{-{e_5}}
   {2\,{\sqrt{2}}} & \frac{-{e_3} - {e_6} + {e_9}}
   {2} & 0 & \frac{{e_1} - {e_2}}{4} & \frac{
      -{e_4}}{2\,{\sqrt{2}}} & \frac{-{e_1} -
     {e_2}}{4} \cr \frac{-{e_7} + {e_8}}
   {4} & \frac{-{e_1} - {e_2}}{4} & \frac{
       -{e_1} + {e_2}}{4} & 0 & \frac{{e_7} +
     {e_8}}{4} & \frac{{e_3} - {e_6} + {e_9}}
   {2} \cr \frac{{e_3} - {e_6} - {e_9}}{2} &
    \frac{{e_5}}{2\,{\sqrt{2}}} & \frac{{e_4}}
   {2\,{\sqrt{2}}} & \frac{-{e_7} - {e_8}}
   {4} & 0 & \frac{-{e_7} + {e_8}}{4} \cr
    \frac{{e_7} + {e_8}}{4} & \frac{-{e_1} +
     {e_2}}{4} & \frac{{e_1} + {e_2}}{4} &
    \frac{-{e_3} + {e_6} - {e_9}}{2} & \frac{{e_
       7} - {e_8}}{4} & 0 \cr  } \right)  \nonumber\\
\label{U6}
\end{eqnarray}
\subsection{The $\mathrm{N^{010}}$ tri-sasakian manifold}
The space $\mathrm{N^{010}}$ can be simply defined as the coset space
\begin{equation}
\label{n010}
{\mathcal{S}\over \mathcal{R}} =
{\mathrm{SU}(3)\over \mathrm{U}(1)}~,
\end{equation}
where, using the Gell-Mann matrices $\lambda^{\overline{A}}$ as
$\su(3)$ generators, the quotient is taken  with respect to
the $\mathrm{U}(1)$ subgroup generated by $\lambda^8$.
The space $\mathrm{N^{010}}$, an instance in the series of 7-dimensional coset
spaces named $N^{p,q,r}$ in the classification of \cite{classificoleo}, is the  only
7-dimensional coset that, when used as a compactification manifold for 11D
supergravity, can preserve $\mathcal{N}=3$ supersymmetry \cite{castn010}. The complete
KK spectrum of the $\mathrm{N^{010}}$ compactification was derived in
\cite{n010massspectrum}, and its $\mathrm{Osp}(3|4)$ multiplet structure
elucidated in \cite{Fre':1999xp,osp34}.
\par
The isotropy group of $\mathrm{N^{010}}$ is $\mathrm{SU}(3)\times \mathrm{SU}(2)$;
the $\mathrm{SU}(2)$ factor is the normalizer of the $\mathrm{U(1)}$ action and,
explicitly, it is generated by $\lambda^{1,2,3}$.
\par
In this case the underlined capital latin indices from the beginning of the
alphabet run on eight values and span the adjoint representation of
the $\su(3)$ Lie algebra.
\par
Let
\begin{equation}
  \Sigma^{\overline{A}}=(\Sigma^\alpha,\Sigma^8)
\label{su3Sigma}
\end{equation}
be the Maurer-Cartan
forms for $\su(3)$, namely let
\begin{equation}
  \Sigma = \ft{\rm i}{2} \, \Sigma^{\overline{A}} \lambda_{\overline{A}} = g^{-1} d
  g\quad ; \quad g\in\mathrm{SU}(3)
\label{Mcformssu3}
\end{equation}
 so that the Maurer Cartan equations
\begin{equation}
  d\Sigma + \Sigma\wedge\Sigma = 0
\label{mcSU3}
\end{equation}
rewritten in the Gell-Mann basis:
\begin{equation}
  d\Sigma^{\overline{A}} \, + \, \ft 12 \, f^{\overline{A}}_{\phantom{A}\overline{BC}} \, \Sigma^{\overline{B}} \,
  \wedge
  \, \Sigma^{\overline{C}} = 0
\label{Gmsu3}
\end{equation}
define the structure constants
$f^{\overline{A}}_{\phantom{A}\overline{BC}}$ of the $\su(3)$ Lie algebra.
The vielbein corresponding to a generic
$\mathrm{SU}(3)\times \mathrm{SU}(2)$-invariant
metric are obtained from the coset vielbein
$\Sigma^\alpha$ ($\alpha=1,\ldots 7$) by rescaling independently the two
groups associated to $\lambda^{\dot\alpha}$ ($\dot\alpha=1,2,3$) and
$\lambda^{\tilde\alpha}$ ($\tilde\alpha=4,5,6,7$). Indeed such a
decomposition is respected both by the $\mathrm{U}(1)$ quotient and by the
$\mathrm{SU}(2)$ action. Thus we have\footnote{Due to a different
choice of structure
constant, our rescaling $\alpha$ is minus twice the one used in
\cite{castellani99}.}:
\begin{equation}
\label{n010viel}
\mathcal{B}^\alpha = (\alpha^{-1} \Sigma^{\dot \alpha},\beta^{-1}
\Sigma^{\tilde \alpha})~.
\end{equation}
The spin connection $\mathcal{B}^{\alpha\beta}$ and the curvature associated to
these vielbein are straightforwardly computed (see
\cite{castellani99}).
\par
The ``standard'' $\mathrm{N^{010}}$ metric is obtained with the following rescalings:
\begin{equation}
\label{standardn010}
\alpha=-4\,\, e~,\hskip 0.5cm \beta=\pm 4\sqrt{2}\,\, e~.
\end{equation}
It preserves $\mathcal{N}=3$ supersymmetry.
It is known \cite{castn010} that, when $\mathrm{N^{010}}$ is realized as the
coset (\ref{n010}), its Killing spinors must actually be \emph{constant}.
With the rescalings (\ref{standardn010}), there
are 3 independent constant spinors $\eta^A$ ($A=1,2,3$)
that satisfy eq. (\ref{kilspineq7dim}), namely
\begin{equation}
\label{n010kilspin}
-{1\over 4} B^{\alpha\beta}_{~~\gamma} \tau_{\alpha\beta}\,\eta^A
 = e\,\tau_\gamma\,\eta^A~.
 \end{equation}
They transform as a \emph{triplet} under the $\mathrm{SU}(2)$
part of the isometry, which therefore truly acquires the role of the R-symmetry
group $\mathrm{SU}(2)_R$ for the 4-dimensional gauged supergravity that arises
from the compactification.
\par
There is a possible solution that differs from (\ref{standardn010})
only by the sign of the rescaling $\alpha$.
While the sign of $\beta$ is irrelevant, because $\beta$ appears quadratically
also in the spin connection, reversing the sign of $\alpha$
amounts to reversing the sign of the spin connection
(or, equivalently, to changing the orientation
of the manifold). This solution with opposite orientation
preserves no supersymmetry.
\par
In the case with preserved $\mathcal{N}=3$ supersymmetry let us calculate
the $\so(8)$ connection as defined by eq.(\ref{so8Ucon1}):
\begin{equation}
\label{buffo}
U^{\mathrm{SO(8)}} \, \equiv \,  -{1\over 4} \mathcal{B}^{\alpha\beta} \tau_{\alpha\beta}\,
 - e\,\tau_\gamma\,\mathcal{\mathcal{B}}^\gamma
 \end{equation}
 We find its explicit expression as an $ 8 \times 8 $ matrix:
\begin{eqnarray}
&\mathbf{U}^{\mathrm{SO(8)}} \, = &\nonumber\\
& \left (
\begin{array}{cc|ccccc|c}
   0 & 0 & 0 & 0 & 0 & 0 & 0 & 0 \cr 0 & 0 & 0 & 0 & 0 & 0 & 0 & 0 \cr
\hline
    0 & 0 &
   0 & 4\,{\mathcal{B}^5} & -4\,{\mathcal{B}^4} & -4\,{\mathcal{B}^7} & 4\,{\mathcal{B}^6} & 0 \cr 0 & 0 & -4\,
   {\mathcal{B}^5} & 0 & \frac{-\left( {\sqrt{3}}\,{\Sigma }^8 \right) }{2} +
   2\,{\mathcal{B}^3} & -2\,{\mathcal{B}^2} & -2\,{\mathcal{B}^1} & 0 \cr 0 & 0 & 4\,{\mathcal{B}^4} & \frac{
      {\sqrt{3}}\,{\Sigma }^8}{2} - 2\,{\mathcal{B}^3} & 0 & 2\,{\mathcal{B}^1} & -2\,
   {\mathcal{B}^2} & 0 \cr 0 & 0 & 4\,{\mathcal{B}^7} & 2\,{\mathcal{B}^2} & -2\,{\mathcal{B}^1} & 0 & \frac{-\left(
        {\sqrt{3}}\,{\Sigma }^8 \right) }{2} - 2\,{\mathcal{B}^3} & 0 \cr 0 & 0 & -4\,
   {\mathcal{B}^6} & 2\,{\mathcal{B}^1} & 2\,{\mathcal{B}^2} & \frac{{\sqrt{3}}\,{\Sigma }^8}{2} +
   2\,{\mathcal{B}^3} & 0 & 0 \cr
\hline
    0 & 0 & 0 & 0 & 0 & 0 & 0 & 0 \cr
   \end{array}
   \right )& \nonumber\\
\label{cbuffo}
\end{eqnarray}
where $\mathcal{B}^\alpha$ is the vielbein defined with the appropriate rescalings already included
and $\Sigma^8$ is the $H$-connection, namely the component along
$\lambda^8$ of the left-invariant $1$-form $\Sigma$ on the coset.
\par
It is visually evident from eq.(\ref{cbuffo}) that the three Killing
spinors are
\begin{equation}
  \begin{array}{ccccccccccc}
    \eta_1 & = & ( 1 & 0 & 0 & 0 & 0 & 0 & 0 & 0 & 0 ) \\
    \eta_2 & = & ( 0 & 1 & 0 & 0 & 0 & 0 & 0 & 0 & 0 ) \\
    \eta_3 & = & ( 0 & 0 & 0 & 0 & 0  & 0 & 0 & 0 & 1 ) \
  \end{array}
\label{trekispi}
\end{equation}
Since the non trivial part of the operator $\mathbf{U}^{\mathrm{SO(8)}}$ is only the
block in the five directions $2,3,4,5,6$.
\par
Hence, in this case,  we have a natural way of introducing an orthogonal basis of sections
of the $\so(8)$ spinor bundle. We use $\eta_{\underline{A}}$ as three basis vectors, while
the other five can be chosen to be
\begin{equation}
  \xi_i = \overrightarrow{\epsilon}_{i+1} \, \quad ; \quad i=1,\dots
  , 5
\label{23456}
\end{equation}
where $\overrightarrow{\epsilon}_i$ are the standard orthonormal
euclidean vectors in eight dimensions.
\par
With this choice the $1$-form connection $\mathbf{U}_{{AB}}$ is just
$8 \times 8$ matrix $\mathbf{U}^{\so(8)}$ as given in eq(\ref{cbuffo}).

\end{document}